%% file: v1.tex
\documentclass[11pt]{article}
\pdfoutput=1
\usepackage{times, amsfonts, amsmath, amssymb, latexsym, setspace, epsfig, hyperref, url, multirow}
\usepackage{graphics, caption, subcaption,scalerel}
\usepackage[dvipsnames]{xcolor}
\usepackage[linesnumbered, ruled]{algorithm2e}

\DeclareMathOperator*{\bigtimes}{\scalerel*{\times}{\sum}}

\def\Pr{{\rm I\!P}}
\def\be{\begin{equation}}
\def\ee{\end{equation}}
\def\bea{\begin{eqnarray*}}
\def\eea{\end{eqnarray*}}
\def\bean{\begin{eqnarray}}
\def\eean{\end{eqnarray}}
\def\nn{\nonumber}
\def\nin{\noindent}

\def\ra{\rightarrow}

\def\Bl{\Bigl}
\def\Br{\Bigr}

\def\R{{\bf R}}

\def\alp{\alpha}

\def\eps{\epsilon}

\newtheorem{Theorem}{Theorem}
\newtheorem{Proposition}{Proposition}

\newtheorem{Lemma}{Lemma}
\newtheorem{Fact}{Fact}

\begin{document}

\setstretch{1.1}

\title{Beta-trees: Multivariate histograms with confidence statements}

\author{Guenther Walther$^1$\thanks{Research supported by NSF grant DMS-1916074} and Qian Zhao$^2$ \\
$^1$Department of Statistics and $^2$Department of Biomedical Data Science\\
 Stanford University\\
\texttt{\{gwalther,qzhao1\}@stanford.edu}}

\date{July 2023}

\maketitle

\begin{abstract}
Multivariate histograms are difficult to construct due to the curse of dimensionality.
Motivated by $k$-d trees in computer science, we show how to construct an efficient data-adaptive partition of
Euclidean space that possesses the following two properties: With high confidence the distribution from which the data
are generated is close to uniform on each rectangle of the partition; and despite the data-dependent construction
we can give guaranteed finite sample simultaneous confidence intervals
 for the probabilities (and hence for the average densities)
 of each rectangle in the partition. This partition will automatically adapt to the sizes of the
regions where the distribution is close to uniform. The methodology 
produces confidence intervals whose widths depend only on the probability content of the rectangles
and not on the dimensionality of the space, thus avoiding the curse of dimensionality. Moreover, 
the widths essentially match the optimal widths in the univariate setting.
The simultaneous validity of the confidence intervals allows to
use this construction, which we call {\sl Beta-trees},
 for various data-analytic purposes. We illustrate this by using Beta-trees for visualizing data and for 
multivariate mode-hunting. 

\end{abstract}

\input{introduction2}
\input{method2}
\input{simulation}

\input{simulation2_mode}
\input{real}
\input{discussion}

\input{proof2}

\bibliographystyle{abbrv}
\bibliography{v1}
\appendix

\input{appendix_kdtree}

\end{document}

%% file: introduction2.tex
\section{Introduction} \label{section:intro}

This paper is concerned with constructing multivariate data summaries for inference.
The classical example of such a data summary is the histogram, which approximates a distribution
with a distribution that is piecewise uniform over rectangles. The two main purposes of a histogram
are to approximate the probability content of subregions and, especially in a lower dimensional setting, 
to visualize  the data. Another important
purpose of the histogram is statistical inference. A relevant example, which is addressed in
some detail in Section~\ref{section:real}, is the analysis
of flow cytometry data. Such data represent multiple parameters of a single cell and
an important  task in the analysis of such data is to detect and isolate subpopulations of cells.
One standard approach to this problem is to construct a multivariate histogram or density estimate
and to identify subpopulations with high density regions~\cite{mckinnon2018fcs}.
However, such a density estimate does not provide any confidence statement about the presence of high density
regions separated by a region of low density. It is a notoriously difficult problem to construct density
estimates in a multivariate setting due to the `curse of dimensionality'; this problem
is compounded by the need to find corresponding standard errors and to adjust for data snooping
when searching for high density regions. In fact, while the vast amount of flow cytometry data
which has become available via modern high-throughput instrumentation has spurred the development
of a large number of algorithms to automate this analysis \cite{DBM, flowcap,brinkman,rahim},
those algorithms generally do not provide any confidence statements about their findings,
and there is a widely recognized need for a principled 
analysis based on statistical guarantees \cite{brinkman,lugli,chatt}.

 Section~\ref{section:real} shows
how the methodology introduced in this paper can be applied to the cytometry problem
to provide guaranteed finite-sample confidence bounds that
allow the detection of such high density regions. A key point is that the widths of
these confidence intervals depend essentially only on the probability content of the region under
consideration and not on the dimensionality of the space, thus avoiding the curse of dimensionality. 
Developing such statistical methodology is crucial for the
quest of the flow cytometry community to discover biological insights by increasing the dimension
of the measurements without having to incur a large penalty in power compared to performing statistical analyses in
a lower dimensional space.

Another example where multivariate density estimates are being used extensively is in databases,
where histograms constitute the most common tool for the succinct approximation of data 
\cite{JKM98,TGIK02,GGI02,ILR12,ADH15}. This is motivated by the fact that often a dataset cannot
be stored in its entirety, so it is necessary to construct a summary (synopsis). Databases typically
summarize data by means of a histogram, and the summary is then used to answer various types of
queries in the same way the original data would have been used \cite{JKM98}. 
Since such a summary of data via a histogram will 
result in some loss of information, it is critically important to provide error bounds (`quality
guarantees') for these histogram estimates. This has lead to a recent active research
effort in the computer science community to derive quality guarantees for 
histograms~\cite{ADH15,ADLS17,CDSS14b,DLS18,DKP18,TGIK02}. The methodology developed
here can be evolved to produce better quality guarantees in that context, and we will report on this
aspect in a different paper.

There are a number of other areas where multivariate histograms play an increasingly important role,
most notably in astronomy and particle physics, see \cite{RamGray2011,scargle, pollack} for surveys.
While there has been an intensive effort in the statistical research community during the last 40 years 
to develop increasingly sophisticated
density estimation methods, the histogram continues
to be surprisingly popular. The following statement
from the astronomical overview paper \cite{scargle} is illuminating as to why the histogram is
a preferred tool in modern astronomical research:

\begin{quote}
"For example, while smoothed plots of pulses within
gamma-ray bursts (GRBs) make pretty pictures, one is really interested in pulse locations, lags,...
All of these quantities can be determined directly from the locations, heights, and widths of the
blocks [of the histogram] - accurately and free of any smoothness assumptions." 
\end{quote}

Implicit in this statement is the claim that an appropriately constructed histogram gives 
a simple summary of the data (in terms of a piecewise constant function) while still allowing to infer
the relevant features of the data. 
Of course, the key point here is that the histogram needs to be constructed appropriately, i.e.
by choosing the partition (the number and location of the bins) appropriately. The definition of the
histogram does not specify these parameters and leaves that choice to the data analyst \cite{freedmanpurves}.
The main contribution of this paper is to provide such a specification which results in favorable
statistical properties. Our method is motivated by $k$-d trees in computer science, which produce an efficient
partition of space that adapts to the data. It turns out that important statistical properties of such $k$-d trees
can be described by the beta distribution. We then show how this fact can be used to prune the $k$-d tree
in a data-adaptive way such the resulting partition has the following two properties: First, with  high confidence the
distribution is close to uniform on each rectangle of the partition. Second, despite the data-dependent construction
we can give guaranteed finite sample simultaneous confidence bounds for the probabilities 
(and hence for the average densities)
 of each rectangle in the partition. These two properties show that the resulting histogram is an appropriate summary
of multivariate data that allows finite sample inference for the tasks described above. 
Moreover, using the multi-scale Bonferroni adjustment of \cite{walther2022scan} results in widths of these confidence 
intervals that do not depend
on the dimension of the space, and furthermore the widths match the optimal widths in a univariate
setting. In that sense the methodology avoids the curse of dimensionality.

$k$-d trees are a popular data structure in computer science that is effective for several important
applications involving multivariate data. In contrast, its statistical properties
in terms of the beta distribution do not yet seem to be available in the literature about $k$-d trees. 
These statistical properties
together with the multi-scale Bonferroni adjustment of \cite{walther2022scan} are the key components
for the proposed methodology, which we therefore call {\sl Beta-tree}.

Before describing our method in Section \ref{section:kd},  we give a review of prior work that is relevant
for the problem considered here. In the univariate setting, \cite{freedman1981hist} and \cite{BirgeRozenholc}
derive rules for choosing the number of bins in a histogram when the bin widths are of equal size.
The common approach is to regard the histogram as a density estimator and to minimize the asymptotic
mean integrated square error, which is of order $n^{-2/3}$. In contrast, in the $d$-dimensional setting
this optimal error is of the order $O(n^{-\frac{2}{2+d}})$, \cite[Section~3.4]{scott2015densitybook}.
Analogous results obtain when employing other standard density estimators, such as the kernel density 
estimator, see \cite{Stone1980}.
These results show that in order to achieve the same mean integrated square error as in the univariate case, 
the number of observations needs to increases \emph{exponentially} with the dimension $d$.
This phenomenon is known as the curse of dimensionality. While the name was introduced by \cite{bellman1961book} in 
connection with computational effort, the statistical version of the curse of dimensionality
 refers to the phenomenon that data become sparse
in high dimensional space. For instance, if one samples $n$ observations from a uniform distribution in a 
$d$-dimensional unit cube, then the number of points in a sub-cube with side length $r$ is about $n r^d$. Hence
in order to obtain the same number of observations in the sub-cube as in the univariate case, the number of
observations needs to increase exponentially with the dimension $d$.

There are several proposals in the literature that involve an adaptive partition of multivariate space, see 
\cite{Ooi2002,Klemela2009,RamGray2011,luo2013bsp,liu2014sieve,liu2015bayesianpartition,li2016star}. 
These proposals use a penalty criterion or maximum likelihood 
in order to select a partition out of a collection of candidate partitions, where the candidate partitions
are obtained from a starting rectangle by recursively subdividing according to Lebesgue measure.
The main statistical results of these papers are rates of convergence for the density estimator resulting from
the adaptive partition. For example, \cite{liu2014sieve} show that when the underlying density can be approximated
well by functions that are piecewise constant on a dyadic partition, then the rate of convergence does not depend on
the dimension of the space. However, none of these proposals provide statistical guarantees such as confidence bounds,
which are required for statistical inference such as the mode-hunting problem addressed below.
A key distinction in our construction is that the recursive partitioning is done according to empirical
measure rather than Lebesgue measure. This allows to obtain finite sample confidence bounds for the probability
content of the resulting rectangles despite the fact that the construction was performed in a data-dependent way.

\subsection{Contributions of his paper}

The prior methods reviewed above assess histogram accuracy in terms of \emph{aggregated} accuracy 
over the entire space, such as Hellinger distance and KL-divergence. The resulting rates of convergence
do not provide any statistical guarantees such as confidence bounds. In this paper, we consider the original purpose
of the histogram: providing  good simultaneous estimates for probabilities over rectangles.   
We construct a concise summary of multivariate data in terms of a histogram such that with high
confidence the distribution is close to uniform on each rectangle in the partition.
We obtain simultaneous confidence intervals for the probabilities of these rectangles that
satisfy finite sample guarantees. We also show that the lengths of these confidence intervals
are close to the optimal lengths in the univariate setting, so this method pays only a very small price
for analyzing multivariate data and therefore avoids the curse of dimensionality. These theoretical results
are derived assuming only that the distribution is continuous.

The Beta-tree histogram can be seen as the statistical counterpart of the $k$-d tree, and it shares the advantageous
property of a compact representation of the data. The Beta-tree prunes subtrees of the $k$-d tree such that the resulting
histogram has fewer regions in the partition while still passing an appropriate goodness-of-fit test.
As in the case of the $k$-d tree, the computational complexity of the Beta-tree is essentially linear in the sample size,
irrespective of the dimension.

Our approach is motivated by the essential histogram of \cite{li2020essential}, which is the
 univariate histogram with the fewest number of 
bins that still passes the generalized likelihood ratio test.
The essential histogram also provides statistical guarantees, but it is not clear how to extend
the essential histogram to the multivariate setting. The key difficulty is to construct an adaptive
partition; furthermore, it is not clear how to carry out the likelihood ratio test on such a data-adaptive partition.
The Beta-tree addresses the first problem by pruning subtrees of the $k$-d tree rather than adding splits to
the partition based on an optimization problem. It addresses the second problem by applying the multiscale
Bonferroni adjustment of \cite{walther2022scan} to the resulting exact beta distributions rather than applying a scale-dependent
penalty on the likelihood ratio statistic. The Beta-tree may therefore have some superfluous splits,
but we submit that there is not much gained by insisting on the minimum number of splits. In turn, the Beta-tree
is much faster to compute than the univariate essential histogram and moreover appears to produce tighter confidence bounds
in the univariate setting.

In the remainder of this article, we will describe our method in Section \ref{section:kd} and illustrate our method 
for data visualization in Section~\ref{section:visualization}, for mode hunting in Section~\ref{section:mode-detection},
and with a real data example in Section \ref{section:real}. In particular, Section \ref{section:mode-detection} 
shows how the simultaneous
inference may be used for multivariate cluster analysis, which may be of independent interest.
Proofs are deferred to Section \ref{section:proof}.

%% file: method2.tex
\section{Constructing the Beta-tree} \label{section:kd}

The construction of the Beta-tree proceeds as follows:
We grow a  $k$-d tree and then on each node (i.e. rectangle) that is bounded, we perform a goodness-of-fit test to decide whether
the data on that rectangle follow a uniform distribution; 
if so, we cut the sub-tree below that
rectangle.

The goodness-of-fit test checks uniformity on a rectangle $R$ by examining the empirical density on
the sub-rectangles of $R$ in the $k$-d tree.
%   The goodness-of-fit test checks whether the empirical density on the rectangle $R$ is contained in the
% intersection of the  confidence intervals for the average densities pertaining to sub-rectangles of $R$ in the $k$-d tree.
This analysis requires to construct
simultaneous confidence intervals for the probability contents of all rectangles
in the $k$-d tree. It turns out that it
is possible to derive such confidence intervals with an exact finite sample level, despite the data-adaptive nature of
the $k$-d tree. Moreover, by applying a certain
weighted multiscale Bonferroni adjustment, it is possible to construct the simultaneous confidence intervals
such that their widths match the optimal univariate widths.

The details for these various items are given in the following subsections.
 
\subsection{Building a space partition with a $k$-d tree}\label{section:construct_kd}

Given $d$-dimensional data $X_i=(X_{i1},\ldots,X_{id}),\ i=1,\ldots,n$, we apply the following recursive
space-partitioning scheme. Apart from some details, this amounts to building a $k$-d tree, see
\cite{bentley1975binarytree,friedman1977}:

We split $R_0:=\R^d$ into two halfspaces by cutting the $p$th coordinate (starting with $p=1$) at an order statistic
of $\{X_{ip}, i=1,\ldots,n\}$ such that (about) half of the observations fall in each
of the resulting two halfspaces:
$$
R_1:= \{ x \in R_0:\ x_p < X_{(\lceil \frac{n}{2} \rceil),p} \},\ \ \ 
R_2:= \{ x \in R_0:\ x_p > X_{(\lceil \frac{n}{2} \rceil),p} \}.
$$
Then we recursively apply this partitioning scheme in $R_1$, using only the observations
that fall into $R_1$ when computing the median, and likewise for $R_2$. Thus the rectangle $R_k$ is split into
two children rectangles $R_{2k+1}$ and $R_{2k+2}$ at a marginal median of the data in $R_k$. 
Note that the observations that determine the boundaries do not belong to any $R_k$. 
We let $p$ cycle through $\{1,\ldots,d\}$ as we progress through the recursion. That is,
we set $p=D \mod d +1$, where $D=\lfloor \log_2 (k+1)\rfloor $ is the
%we set $p=D \mod d$ if $D \mod d \neq 0$ and $p=d$ otherwise, where $D=\lfloor \log_2 (k+1)\rfloor +1$ is the recursion
tree depth of $R_k$.
%\be \label{1}
%p = \begin{cases} 
%D \mod\ d \quad & \text{if} \ D \mod\ d \neq 0, \\
%d\quad & \text{otherwise}. \\
%\end{cases}
%\ee
We stop splitting $R_k$ once it has fewer than $4 \log n$ observations. 

Some rectangles $R_k$ in the $k$-d tree will be unbounded, whereas the construction of a histogram is necessarily 
resticted to bounded rectangles. The following modification produces a $k$-d tree with only bounded rectangles,
which allows to extend the construction of a histogram 
further out into the tails of the data:
We create  a bounding box by
discarding the observations with the smallest and with the largest order statistic in the first coordinate,
and we use these two order statistics (or some other order statistics if one wishes to cut a larger fraction
of the observations) 
as bounding values in the first coordinate. We iterate this process
through all $d$ coordinates. Then we run the above space-partitioning algorithm on the remaining observations
with $R_0$ equaling the bounding box. Importantly, all of the statistical methodology described below continues to
hold with this modification, in particular the crucial finite sample result given in Proposition~\ref{beta1}.
 
If the data $X_{1p},\ldots,X_{np}$ are distinct for each coordinate $p$, then the number of
observations $n_k$ in $R_k$ is a function of $n$ and $k$ only.
It is convenient to define the empirical measure as $F_n(R_k)=\frac{n_k+1}{n}$
rather than $\frac{n_k}{n}$. 
We discuss this as well as other relevant aspects of  
$k$-d trees in Appendix \ref{appendix:kdtree}. 

\subsection{Deriving exact confidence bounds for the rectangles in the $k$-d tree}

We denote the distribution of the $X_i$ by $F$. The following proposition is the starting point
for our construction. It shows that $F(R_k)$
follows a beta distribution whose parameters depend only on $n_k$ and the sample size $n$: 

\begin{Proposition} \label{beta1}
Let $X_i$, $i=1,\ldots,n$, be i.i.d. $F$, where $F$ is a continuous distribution
on $\R^d$. Then every rectangle $R_k$ generated by the partitioning scheme
in Section~\ref{section:construct_kd} is a random set containing a deterministic
number $n_k$ of observations in its interior and the random variable $F(R_k)$ satisfies
$$
F(R_k) \sim \textrm{Beta}\, (n_k+1,n-n_k). 
$$
\end{Proposition}

This result is related to early results by Wald~(1943) and Tukey~(1947), see the comments in the proof 
in Section~\ref{section:proof}.
We note that Proposition~\ref{beta1}  remains valid for other ways of choosing the axis $p$ for
the split, e.g. choosing $p$ randomly, as well as for other ways to set $n_k$,
as long as these do not depend on the data $\{X_i\}_{i=1}^n$. 

An important consequence of Proposition~\ref{beta1} is that $F(R_k)$ is a pivotal quantity, i.e. its distribution does not
depend of $F$. Furthermore, this distribution is known exactly.
This is a multivariate generalization of the
well known fact that in the univariate case $F((X_{(i)},X_{(j)})) \sim \textrm{Beta }
(j-i,n+1-(j-i))$, see Shorack and Wellner~(1986) \cite[Chapter~3.1]{Shorack1986EmpiricalProcess}.
We note that this property depends crucially on employing
the data adaptive collection $\{R_k\}$ rather than constructing a partition by splitting according to
Euclidean distance as is usually done in the literature.

Proposition~\ref{beta1} implies an exact
 $(1-\alpha)$ level confidence interval for $F(R_k)$:
\begin{equation}\label{eq:ci_rk}
C_k(\alp):=\ \left(q\mathrm{Beta}(\frac{\alpha}{2}, n_k+1,n-n_k), \,
q\mathrm{Beta}(1-\frac{\alpha}{2}, n_k+1,n-n_k)\right)
\end{equation}
where $q\mathrm{Beta}(\alpha,\cdot,\cdot)$ denotes the $\alpha$-quantile of the beta distribution with the
given degrees of freedom.
Strictly speaking, this is a prediction interval or a tolerance region
since $F(R_k)$ is a random variable, which measures the probability content of the random set $R_k$. 
Likewise, an exact $(1-\alpha)$ level confidence interval for
the average density $f(R_k) = F(R_k)/|R_k|$  is given by dividing the bounds in (\ref{eq:ci_rk})
by the volume $|R_k|$, see (\ref{3}) below.

\subsection{Constructing  simultaneous confidence bounds with a multiscale Bonferroni adjustment}

In order to construct confidence intervals for the $F(R_k)$ and $f(R_k) = F(R_k)/|R_k|$ that are simultaneously valid for all 
rectangles $R_k$ in the $k$-d tree, we use the weighted Bonferroni adjustment that 
\cite{walther2022scan} propose for univariate
scan statistics. The motivation for using that weighted adjustment is to obtain good statistical performance
across all scales, which in this context means across all tree depths $D$. The prescription given in \cite{walther2022scan}
is to assign the same significance level to each interval at a given scale, and to weigh the significance level
across scales according to a harmonic sequence so that the smallest scale is weighted with a factor $\frac{1}{2}$, the second
smallest scale with a factor $\frac{1}{3}$ etc. Translated to the setting of a $k$-d tree with $R_0$ being a bounding box, 
this prescription assigns
each of the $N_D$ bounded rectangles at tree depth $D$ the significance level
\be \label{eq:weighted_bonferroni}
\alpha_D = \frac{\alp}{N_D (D_{max}-D+2) \sum_{B=2}^{D_{max}+1} \frac{1}{B}},\ \ D\geq 1,
\ee
where $D_{max}$ is the maximum depth of the Beta-tree, and $\alpha_0=0$.
%\be \label{eq:weighted_bonferroni}
%\alpha_D = \frac{\alp}{N_D (D_{max}-D+2) \sum_{B=2}^{D_{max}-2} \frac{1}{B}},\ \ D>3.
%\ee
%where $D_{max}$ is the maximum depth of the Beta-tree. We compute these $\alpha_D$ only for
% depths $D>3$ as smaller depths are typically not relevant for the goodness-of-fit test described
%below, so we set $\alpha_D:=0$ for $D \leq 3$.
Therefore, if $R_k$ has tree depth $D$, then

\begin{align}  \label{3}
\mathrm{lower}(R_k)  =\frac{q\textrm{Beta }(\frac{\alp_D}{2},n_k+1,n-n_k)}{|R_k|}, \quad 
\mathrm{upper}(R_k)  = \frac{q\textrm{Beta }(1-\frac{\alp_D}{2},n_k+1,n-n_k)}{|R_k|},
 \end{align}

provide lower and upper confidence bounds for $f(R_k) = F(R_k)/|R_k|$, and these confidence bounds have
simultaneous coverage level $1-\alpha$ for all bounded $R_k$ since the corresponding $\alp_D$
sum to $\alpha$.

Likewise, using $C_k(\alp_D)$ in (\ref{eq:ci_rk}) gives simultaneous confidence bounds for the $F(R_k)$. 

If $R_0$ equals $\R^d$ rather than a bounding box, then there will be no bounded rectangles at the smallest depths $D$
and (\ref{eq:weighted_bonferroni}) changes accordingly. Appendix~\ref{appendix:kdtree} gives the details.

\subsection{Pruning the $k$-d tree by checking goodness-of-fit}

A key principle for constructing a histogram is to find a parsimonious representation which still gives
an good approximation to the distribution, see e.g. the discussion in \cite{li2020essential}. This principle can be readily
implemented for a nested partition as follows: We keep the largest rectangles $R_k$ for which we are confident that the 
distribution of the data  on $R_k$ is the uniform distribution  specified by a histogram that uses $R_k$
as a bin\footnote{More precisely, the GOF test is not about the conditional distribution on $R_k$ but about the
unconditional $F$ restricted to $R_k$}.
Assessing whether the distribution is uniform on a rectangle $R_k$ amounts to a goodness-of-fit test. 
Such a test can be readily implemented
for the multiscale partition given by the $k$-d tree since the rectangles $R \subset R_k$ constitute an appropriate
collection of test sets for which we can compare the empirical density to that on $R_k$. This simply amounts
to checking whether the empirical density on $R_k$ lies in all the confidence intervals (\ref{3}) for the 
$f(R)$, $R \subset R_k$,
i.e. in the intersection of these confidence intervals. This intersection is given by $(\widetilde{\mathrm{lower}}(R_k),
\widetilde{\mathrm{upper}}(R_k))$, where these bounds are defined recursively as follows:

\begin{align} \label{eq:ci_tilde}
\widetilde{\mathrm{lower}}(R_k) & = \max \left(\mathrm{lower}(R_k),
\,\widetilde{\mathrm{lower}}(1st \,\mathrm{child\,of}\,R_k), \,\widetilde{\mathrm{lower}}(2nd \,\mathrm{child\,of}\,R_k)\right) \\
\widetilde{\mathrm{upper}}(R_k) & =\min \left(\mathrm{upper}(R_k),
 \,\widetilde{\mathrm{upper}}(1st\, \mathrm{child\,of}\,R_k), \,\widetilde{\mathrm{upper}}(2nd\, \mathrm{child\,of}\,R_k)\right) \nn 
\end{align}

if $R_k$ has children, otherwise $\widetilde{\mathrm{lower}}(R_k) = \mathrm{lower}(R_k)$ and
$\widetilde{\mathrm{upper}}(R_k) = \mathrm{upper}(R_k)$.

Now we can define the {\sl Beta-tree} as the collection of the maximal (with respect to inclusion) bounded
rectangles of the $k$-d tree
that pass the goodness-of-fit test. That is, we take all rectangles $R_k$ that are bounded and satisfy 
$\widetilde{\mathrm{lower}}(R_k) \leq h_k \leq \widetilde{\mathrm{upper}}(R_k)$ while none of the ancestors
of $R_k$ satisfy these two conditions.
Here $h_k$ is the empirical average density of $R_k$: 
\be \label{6}
h_k := \frac{F_n(R_k)}{|R_k|} = \frac{n_k+1}{n|R_k|}
\ee
The tree structure makes it easy to find these maximal rectangles recursively or iteratively, 
see Appendix~\ref{appendix:kdtree}.

The {\sl Beta-tree histogram} is the histogram constructed using the rectangles in the Beta-tree.
That is, on the rectangle $R_k$ the histogram has height $h_k$ given by (\ref{6}). The Beta-tree histogram
lies in the $(1-\alpha)$ confidence set for $F$ that is given by the multiscale goodness-of-fit test, and it is
the most parsimonious distribution in that confidence set among histograms that use rectangles in the $k$-d tree
as potential bins.

\section{The simultaneous confidence intervals attain the optimal univariate widths}

The Beta-tree satisfies a key goal of the histogram: It provides a parsimonious summary of the data while still
giving a good approximation to the distribution. Importantly, the data-adaptive construction using the goodness-of-fit
test described in the
previous section does not invalidate the statistical guarantees (\ref{3}) for $f(R_k)$,
since those confidence bounds are simultaneous for all $R_k$.
This raises the question whether this simultaneity results in overly conservative confidence bounds.
It turns out that this is not the case: In fact, the widths of the confidence intervals
essentially match the optimal simultaneous widths in the {\it univariate} setting. 
This shows that this data-adaptive construction
avoids the curse of dimensionality, 
and there is only an asymptotically  negligible price to pay 
compared to the univariate setting for effectively summarizing multivariate data with a histogram.

To make this precise, we first summarize the relevant lower bound in the univariate case. It is well known
that the empirical measure of an interval $I$, $F_n(I)$, estimates $F(I)$ with precision
$\sqrt{n}\frac{|F_n(I)-F(I)|}{\sqrt{F(I)(1-F(I))}}=O_p(1)$. If one wishes to estimate $F(I)$ simultaneously
for all intervals $I$, then there is an unavoidable penalty of size $\sqrt{2 \log \frac{e}{F(I)}}$:
Theorem~1 in \cite{li2020essential} shows that if ${\cal J}_n=\bigcup_i I_{i,n}$ is a partition of the line
such that $F(I_{i,n})=p_n$, $i=1,\ldots,\lfloor \frac{1}{p_n} \rfloor$ and $\frac{\log^2 n}{n} \leq p_n \rightarrow 0$,
then
\be \label{lower1}
\Pr_F \Bl(\mbox{for some } I \in {\cal J}_n: \sqrt{n}\frac{|F(I)-F_n(I)|}{\sqrt{F(I)(1-F(I))}} \ \ge \ 
\left( \sqrt{2}-\eps_n \right) \sqrt{ \log \frac{e}{F(I)}} \Br)\ \rightarrow 1 \qquad (n\ra \infty)
\ee
with $\eps_n \ra 0$ at a certain rate. Moreover, this penalty cannot be improved with any other estimator in place
of $F_n$. The constant $\sqrt{2}$ is important as it measures the difficulty of the estimation problem, see \cite{walther2022scan}.
The lower bound (\ref{lower1}) implies a lower bound for any confidence interval $C$ for $F(I)$, because if 
$|F(I)-F_n(I)|$ satisfies a lower bound, then the radius $\sup_{G\in C} |G-F_n(I)|$ must also satisfy this bound.

%Note that the rectangles $R_k$ of the $k$-d tree at depth $D$ constitute a partition 
%of $R_0$ with $F(R_k) \approx 2^{-D}$.
Theorem~\ref{optimal} shows that despite the data-adaptive construction in multivariate space, the simultaneous confidence 
intervals $C_k(\alp_D)$ attain the critical constant $\sqrt{2}$ of the univariate lower bound
(\ref{lower1}), up to a term $\eps_n \ra 0$:

\begin{Theorem} \label{optimal}
If $F$ is a continuous distribution on $\R^d$, then for every $k$ with $n_k \in [\log^2 n,n^q]$, $q \in (0,1)$,
the confidence interval $C_k(\alp_D)$ for $F(R_k)$ satisfies
$$
\sup_{G \in C_k(\alp_D)} \sqrt{n} \frac{|G-F_n(R_k)|}{\sqrt{G(1-G)}}\ \leq \ \left(\sqrt{2} +\frac{4}{\sqrt{\log n}} \right)
\sqrt{ \log \frac{e}{G}}
$$
\end{Theorem}

Some remarks:

1. Note that the theorem applies to all confidence intervals in the $k$-d tree, not just those of the pruned Beta-tree. 

2. The inequality in the theorem is deterministic: While $R_k$ is random, $F_n(R_k)=\frac{n_k+1}{n}$
as well as $C_k(\alp)$ are deterministic. In particular, the above inequality holds uniformly in $k$. 

3. The theorem applies to rectangles $R_k$ that are not very large, i.e. $n_k \leq n^q$. 
A different Bonferroni weighting would allow to extend the theorem to all rectangle sizes, but the discussion
in \cite{walther2022scan} for the univariate regression setting suggests that it is worthwhile to trade off the optimality
for large rectangles in order to obtain a  better finite sample performance for smaller rectangles, 
which are typically more relevant in practice.

%% file: simulation.tex
\section{Summarizing data using the Beta-tree histogram}\label{section:visualization}

\begin{figure}[h]
     \centering
     \begin{subfigure}[b]{0.45\textwidth}
         \centering
         \includegraphics[width=\textwidth]{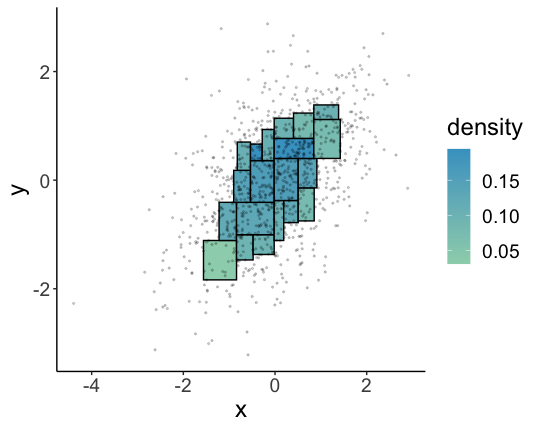}
     \end{subfigure}
       \hfill
     \begin{subfigure}[b]{0.45\textwidth}
         \centering
         \includegraphics[width=1\textwidth]{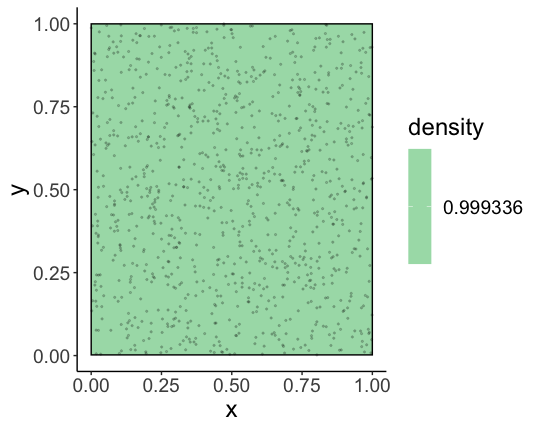}
     \end{subfigure}

        \caption{Beta-tree histograms for samples from a bivariate normal (left) and a bivariate
uniform (right), $n=1000$.}
        \label{fig:normaluniform}
\end{figure}

In this section we apply the Beta-tree histogram to summarize two- and three-
dimensional data from various distributions. The left plot in Figure~\ref{fig:normaluniform} shows
the Beta-tree for a sample from a two-dimensional normal distribution with correlation
coefficient 0.5, while right plot shows the Beta-tree with a bounding box for a bivariate uniform
distribution. Both samples have sizes $n=1000$ and the bounding box is constructed as described
in Section~\ref{section:construct_kd} by exluding 0.5\%
of the data in each tail of each coordinate.

\begin{figure}[h]
     \centering
     \begin{subfigure}[b]{0.45\textwidth}
         \centering
         \includegraphics[width=\textwidth]{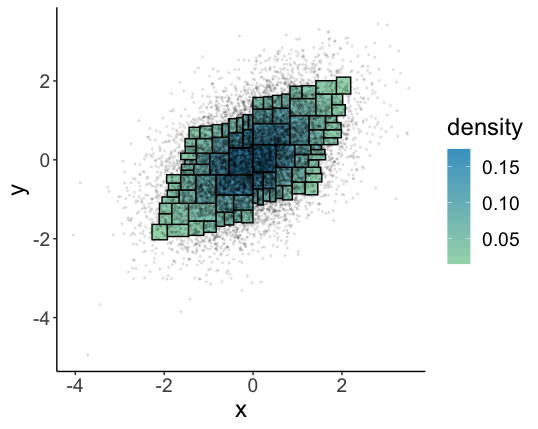}
     \end{subfigure}
       \hfill
     \begin{subfigure}[b]{0.45\textwidth}
         \centering
         \includegraphics[width=1\textwidth]{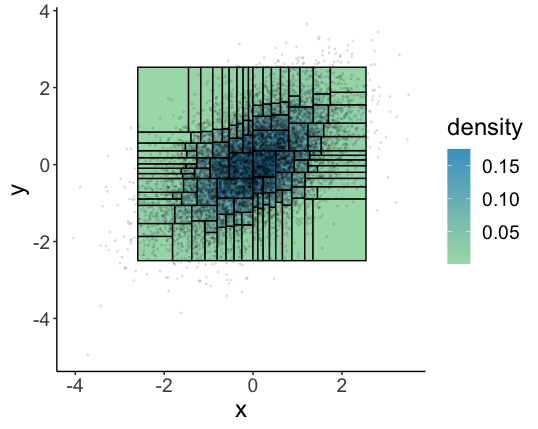}
     \end{subfigure}

        \caption{Beta-tree histograms without a bounding box (left) and with a bounding box (right), $n=20000$.}
        \label{fig:normal}
\end{figure}

Figure~\ref{fig:normal} shows the Beta-tree histogram with and without bounding box for a larger sample of
size $n=20000$ from the same bivariate normal distribution. Due to the construction via the multiscale
goodness-of-fit test, the Beta-tree histogram produces larger bins
where the density does not change much and smaller bins where the density changes quickly.
This is an important and necessary feature in order to obtain a summary that is both parsimonious and accurate.
In particular, the uniform distribution in Figure~\ref{fig:normaluniform} results in a single
bin for the Beta-tree. This desirable outcome is notoriously difficult to achieve
with other histogram rules.

In order to evaluate the Beta-tree for more complex distributions we consider two- and three-
dimensional data from  mixtures of multivariate Gaussian distributions. We consider the following two scenarios:

In the first scenario, we sample $n=2000$ observations from the following two-dimensional mixture:
\[
\frac{2}{5} \mathcal{N}\left(\left(\begin{matrix}-1.5 \\0.6\end{matrix}\right),\left(\begin{matrix}1 & 0.5 \\ 
0.5 & 1\end{matrix}\right) \right) + \frac{3}{5} \mathcal{N}\left(\left(\begin{matrix}2 \\-1.5\end{matrix}\right),
\left(\begin{matrix}1 & 0 \\ 0 & 1\end{matrix}\right) \right). 
\]
In the second scenario, we sample $n = 20,000$ observations from the three-dimensional mixture
\begin{multline*}
\frac{2}{5} \mathcal{N}\left(\left(\begin{matrix}-1.5 \\0.6\\1\end{matrix}\right),\left(\begin{matrix}1 & 0.5 & 0.5 \\ 
0.5 & 1 & 0.5 \\ 0.5 & 0.5 & 1
\end{matrix}\right) \right) + \frac{2}{5} \mathcal{N}\left(\left(\begin{matrix}2 \\-1.5\\0\end{matrix}\right),
\left(\begin{matrix}1 & 0 & 0 \\ 0 & 1 & 0 \\ 0 & 0 & 1\end{matrix}\right) \right) \\
+ \frac{1}{5} \mathcal{N}\left(\left(\begin{matrix}-2.6\\-3\\-2\end{matrix}\right),\left(\begin{matrix}1 & -0.4 & 0.6 \\ 
-0.4 & 1 & 0 \\ 0.6 & 0 & 1\end{matrix}\right) \right). 
\end{multline*}
We chose these two distributions because some of the components are
correlated and because individual components are relatively disjoint from
each other. For example, if we were to classify which component an observation is sampled from using a Bayes 
classifier, then the accuracy is over 98\% in both scenarios. 

\begin{figure}
     \centering
     \begin{subfigure}[b]{0.45\textwidth}
         \centering
         \includegraphics[width=\textwidth]{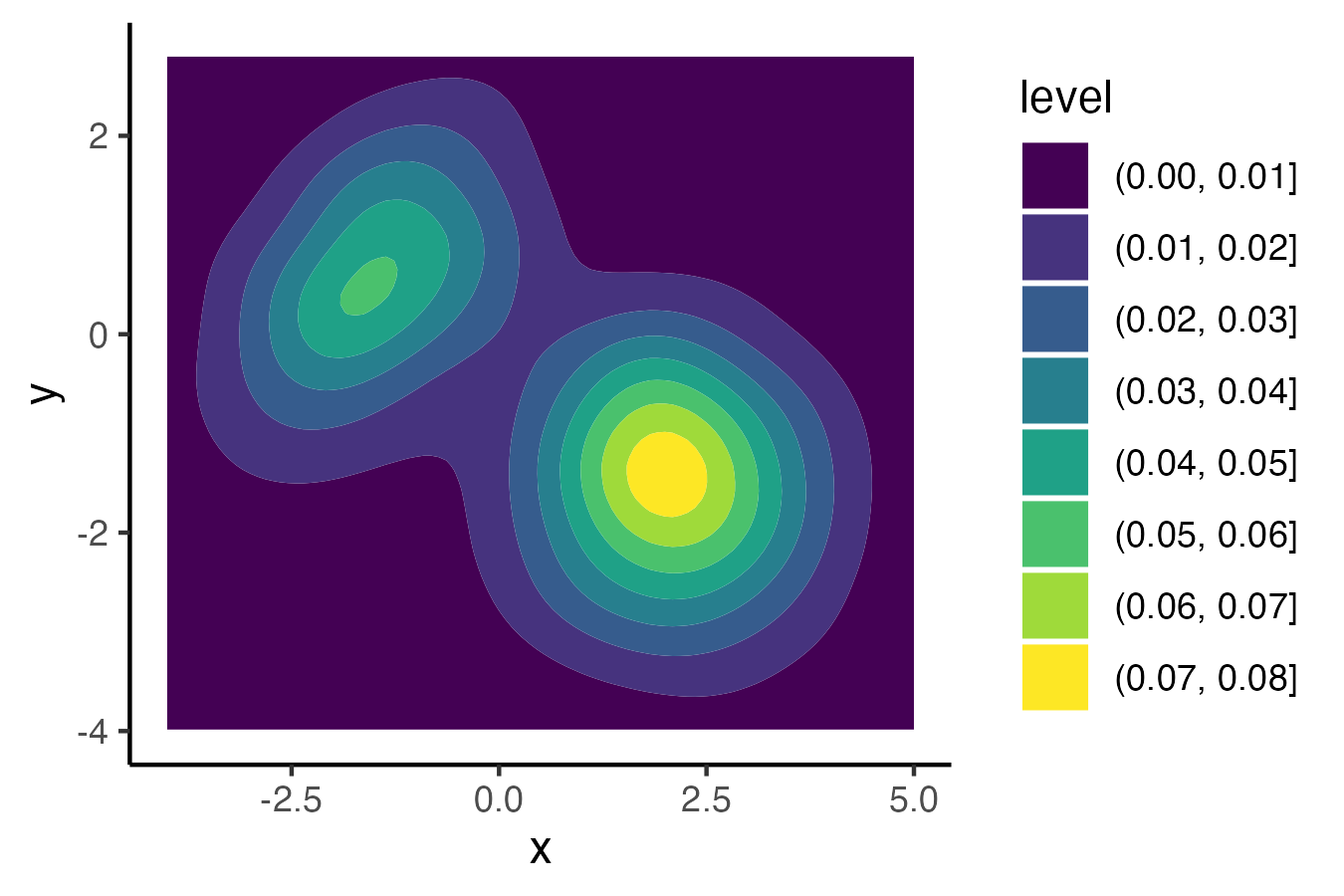}
         \caption{Kernel density estimate}
         \label{fig:kerned2d}
     \end{subfigure}
       \hfill
     \begin{subfigure}[b]{0.45\textwidth}
         \centering
         \includegraphics[width=1\textwidth]{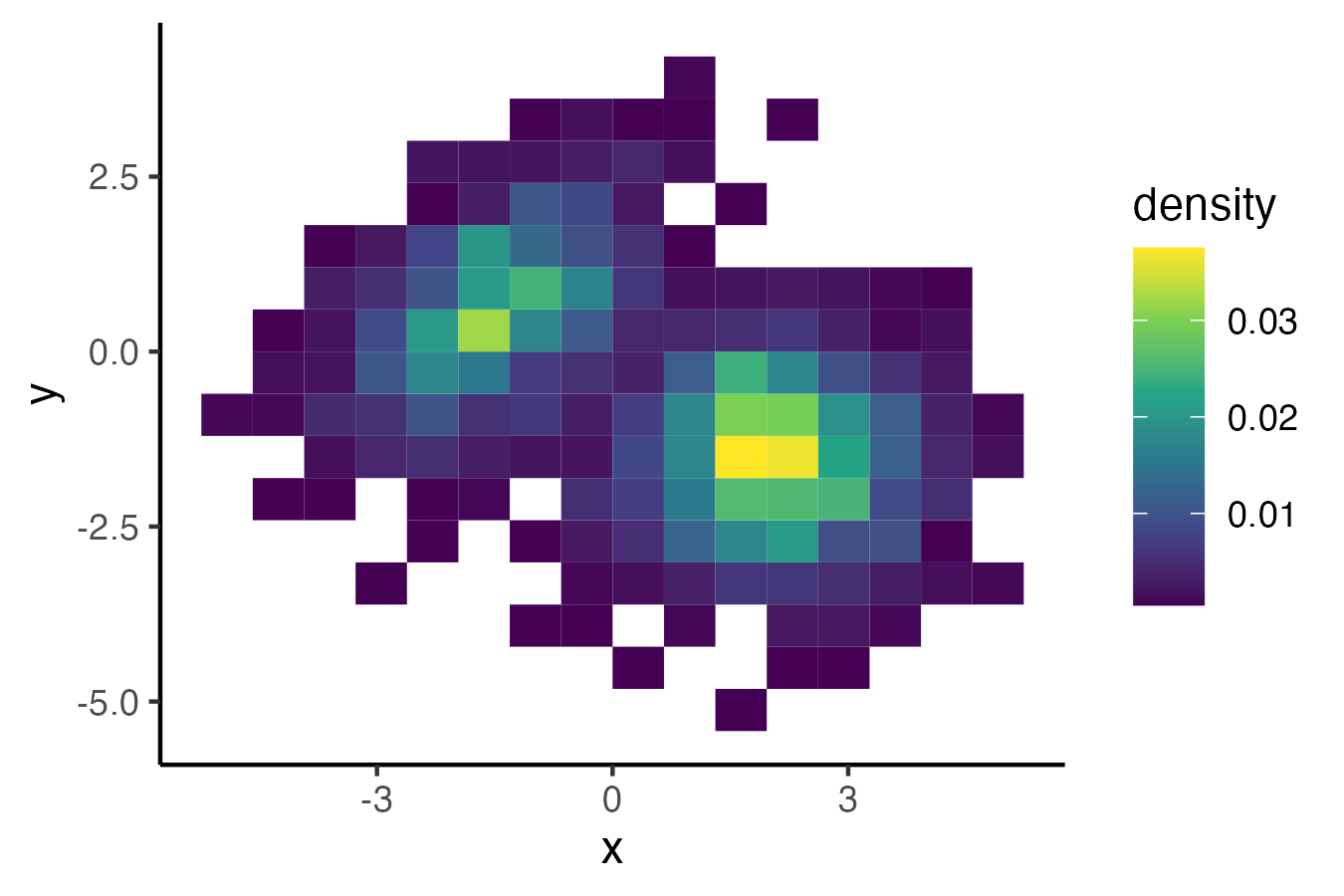}
         \caption{Histogram with 15$\times$15 bins} 
         \label{fig:fixed2d}
     \end{subfigure}
 
     \bigskip
     \begin{subfigure}[b]{0.45\textwidth}
         \centering
         \includegraphics[width=1\textwidth]{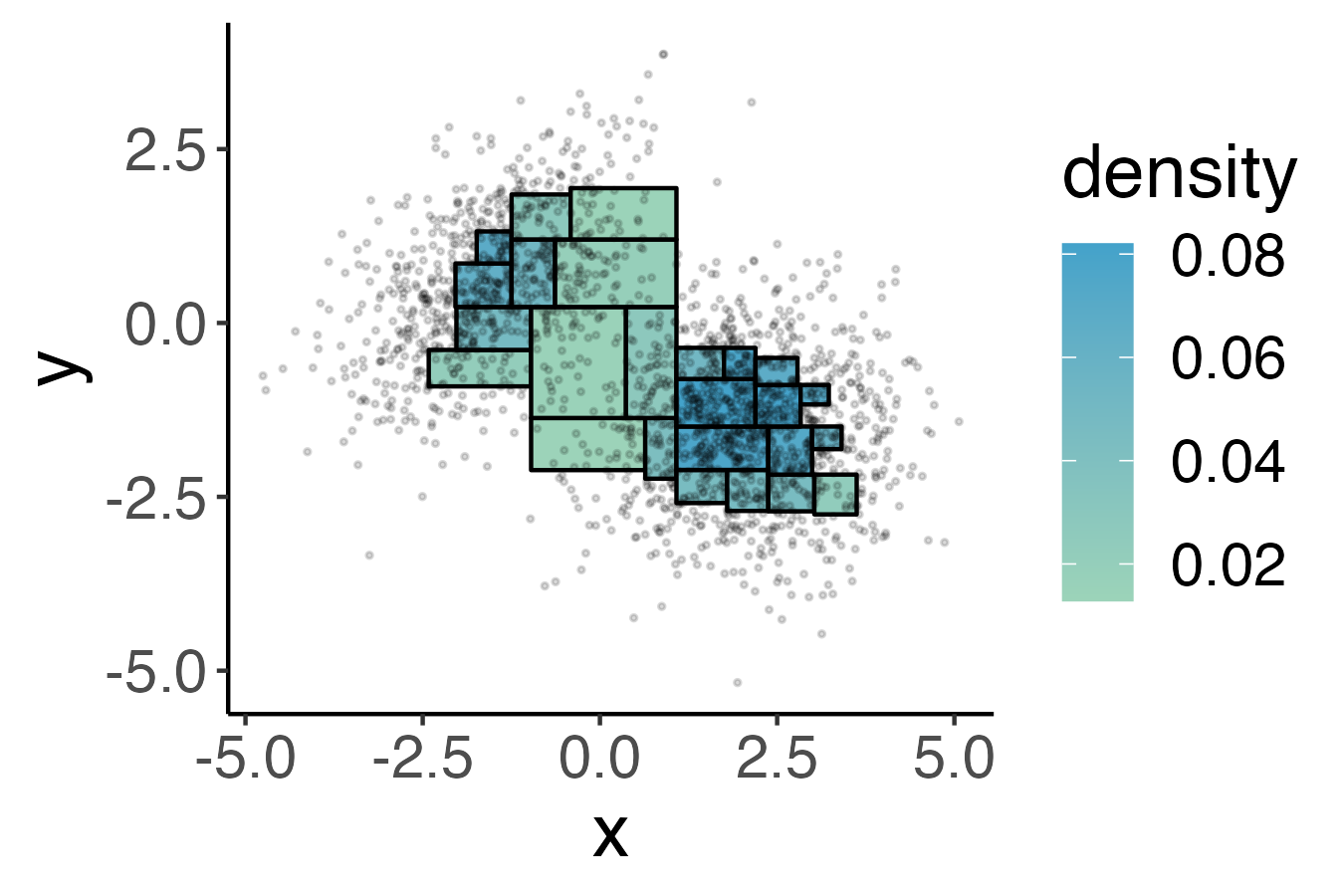}
         \caption{Beta-tree histogram}
         \label{fig:hist2d}
     \end{subfigure}
         \hfill
     \begin{subfigure}[b]{0.45\textwidth}
         \centering
         \includegraphics[width=1\textwidth]{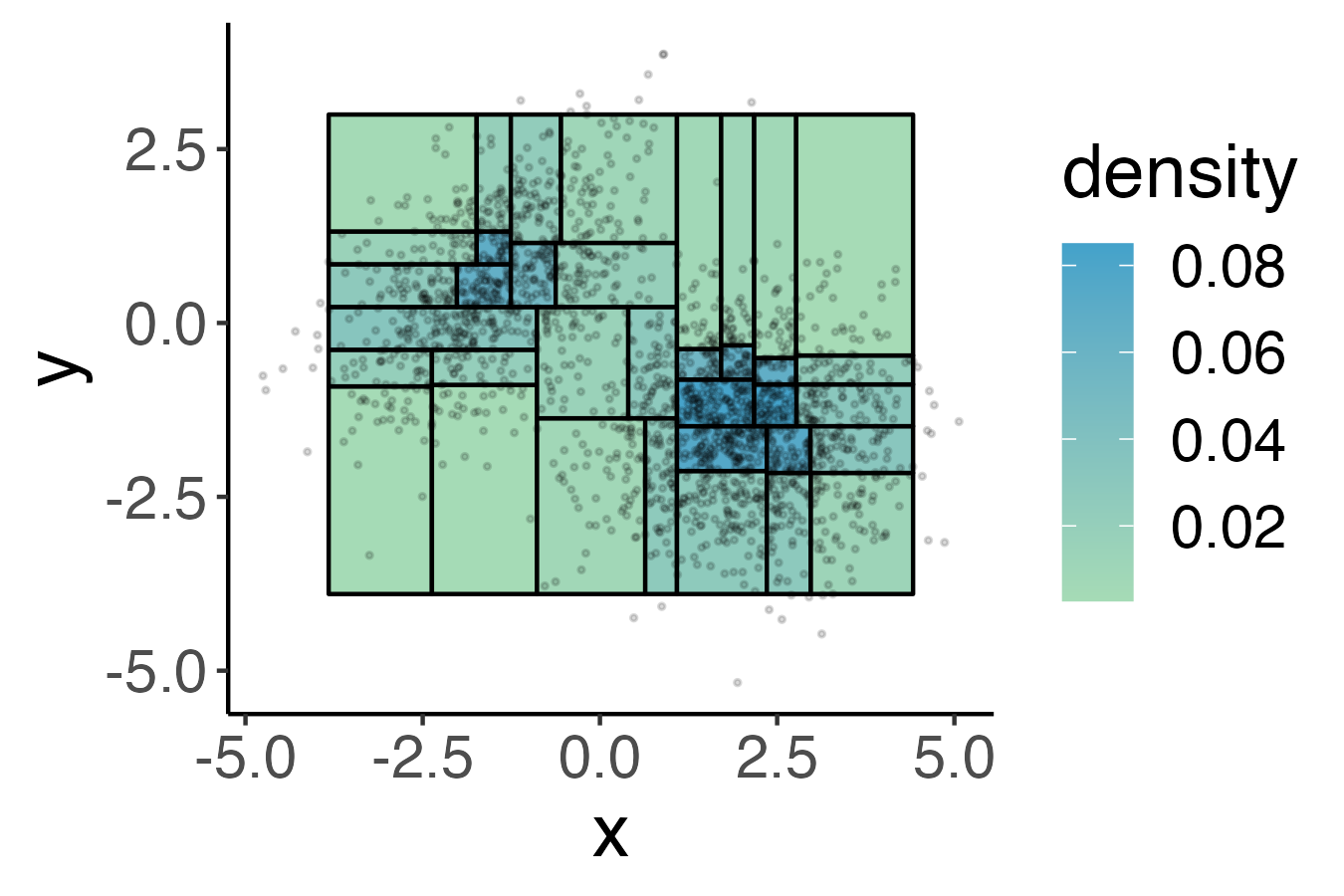}
         \caption{Beta-tree histogram with bounding box}
         \label{fig:bounded2d}
     \end{subfigure}

        \caption{Density estimate and histograms for a mixture of two-dimensional Gaussians, $n=2000$.}
        \label{fig:mixture2d}
\end{figure}

\begin{figure}
     \centering
     \begin{subfigure}[b]{0.45\textwidth}
         \centering
         \includegraphics[width=\textwidth]{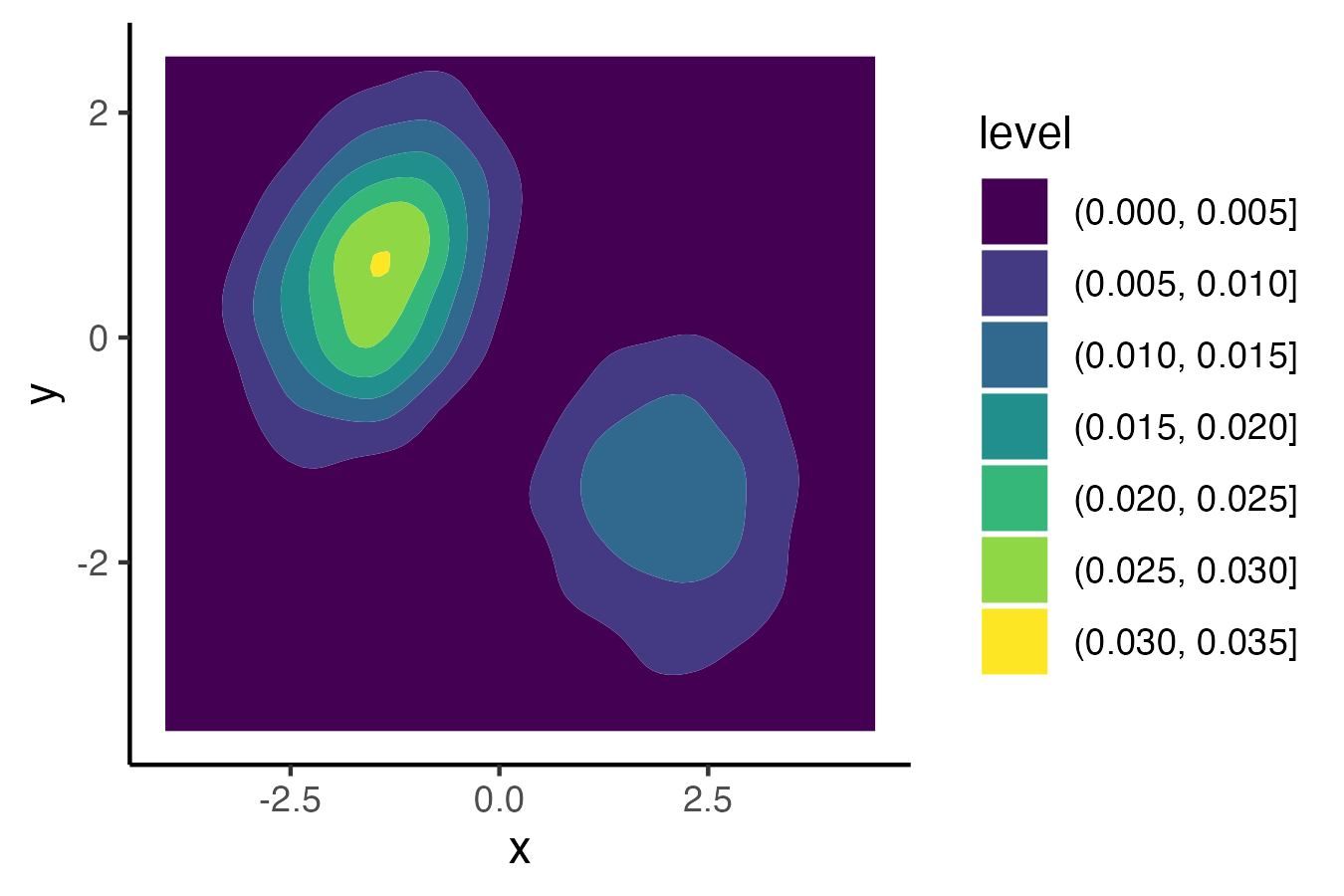}
         \caption{Kernel density estimate}
         \label{fig:kerned3d}
     \end{subfigure}
       \hfill
     \begin{subfigure}[b]{0.45\textwidth}
         \centering
         \includegraphics[width=1\textwidth]{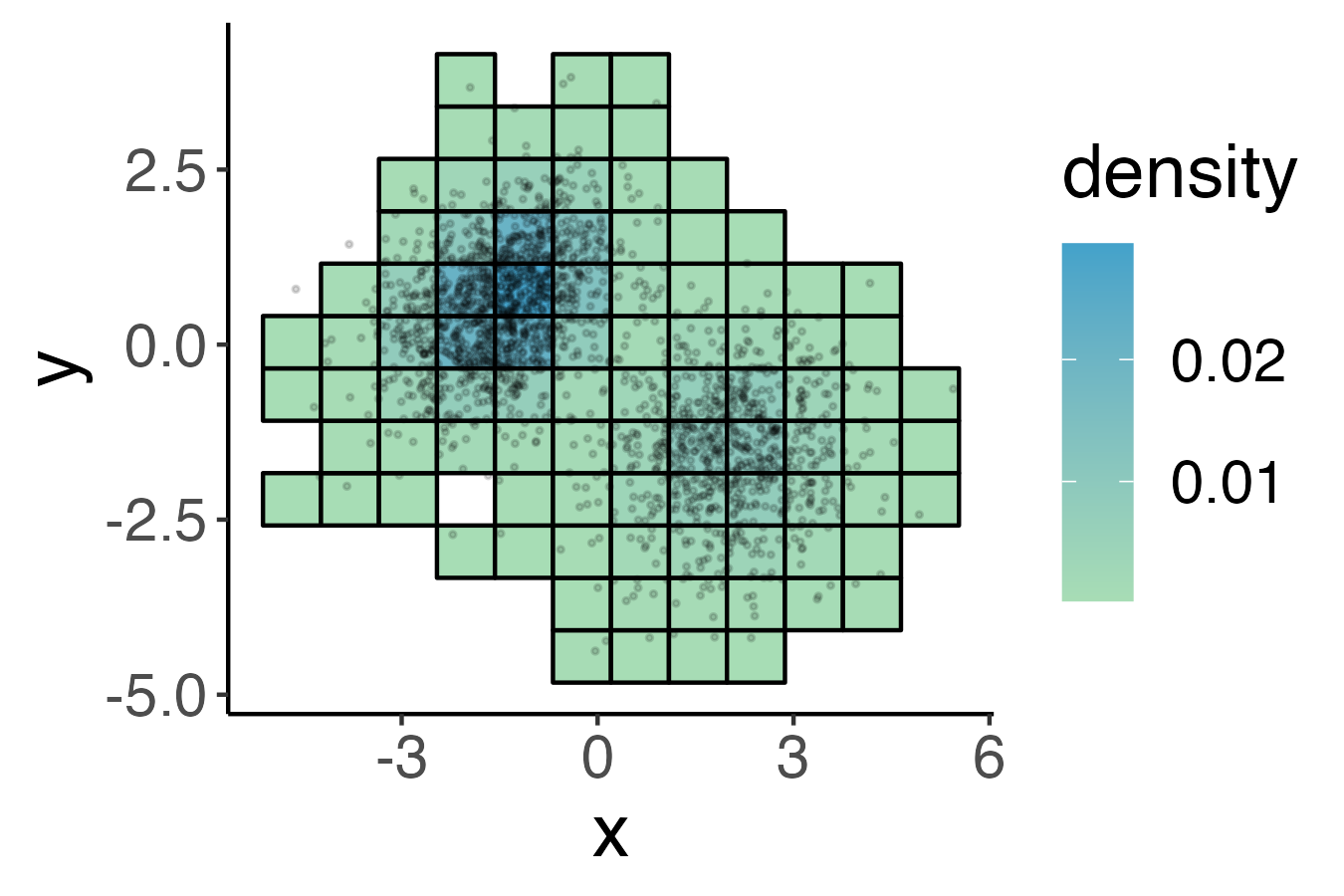}
         \caption{Histogram with 15$\times$15$\times$15 bins }
         \label{fig:fixed3d}
     \end{subfigure}

       \bigskip
     \begin{subfigure}[b]{0.45\textwidth}
         \centering
         \includegraphics[width=1\textwidth]{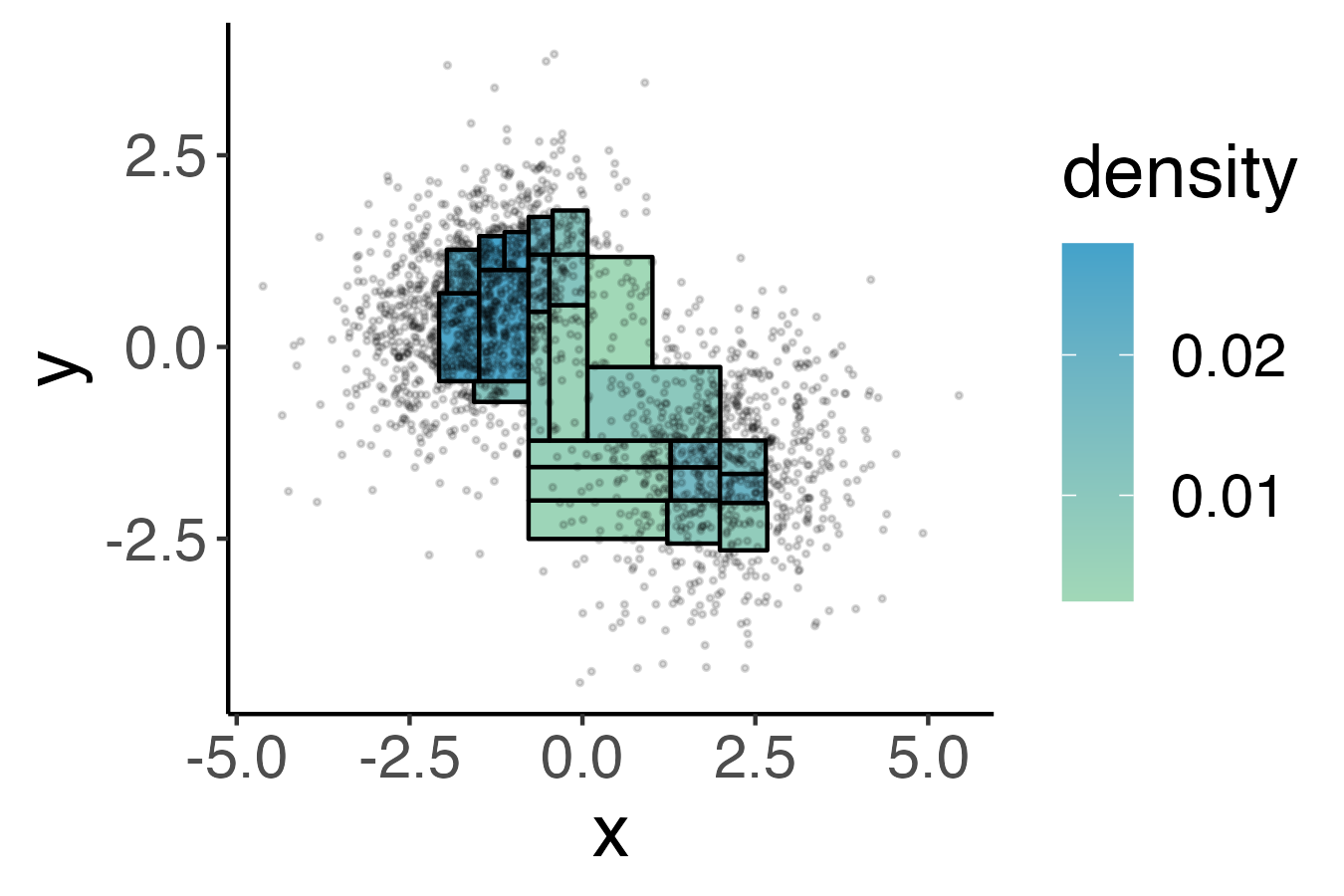}
         \caption{Beta-tree histogram}
         \label{fig:hist3d}
     \end{subfigure}
       \hfill
     \begin{subfigure}[b]{0.45\textwidth}
         \centering
         \includegraphics[width=1\textwidth]{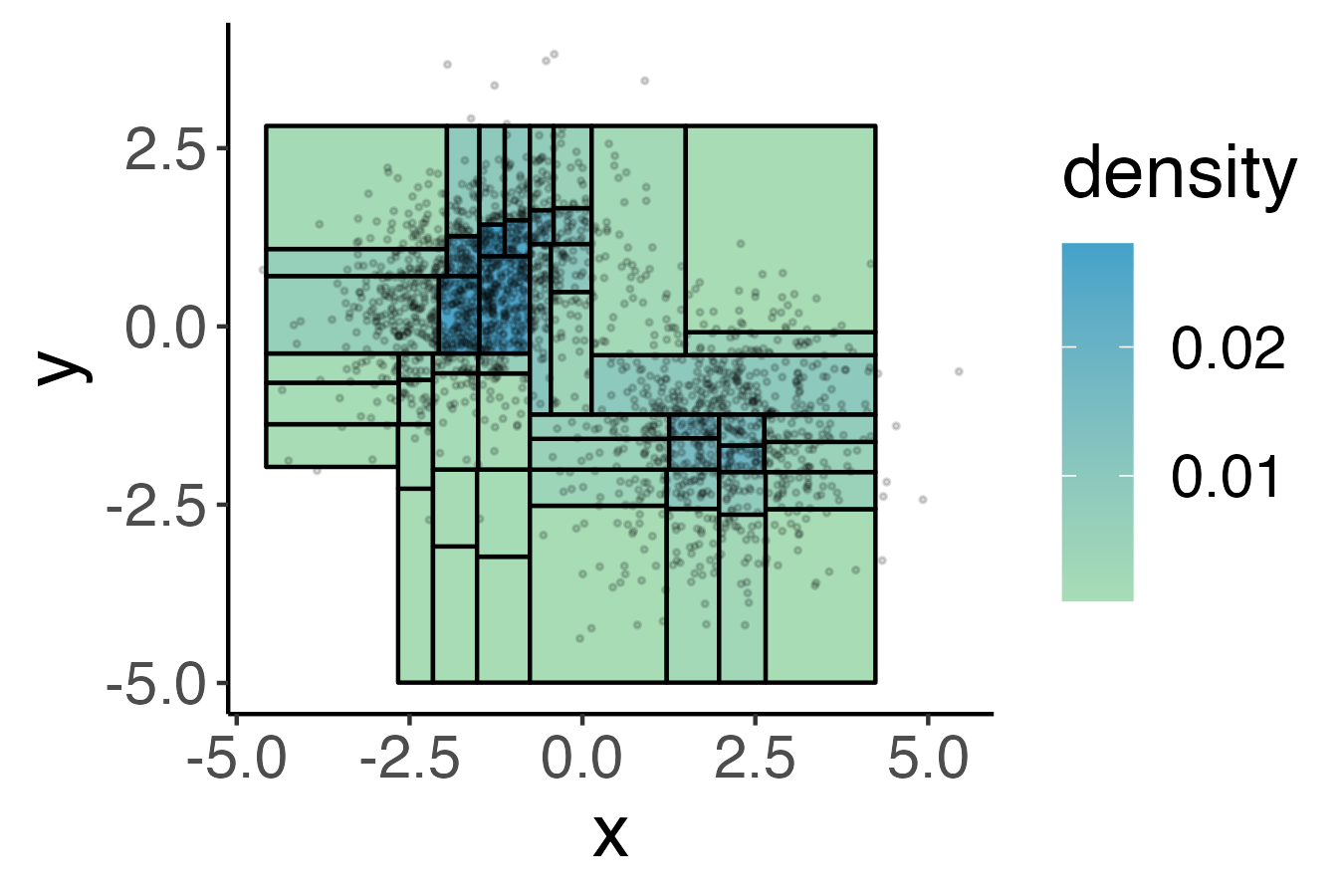}
         \caption{Beta-tree histogram with bounding box}
         \label{fig:bounded3d}
     \end{subfigure}

        \caption{Density estimate and histograms for
 a mixture of three-dimensional Gaussians, $n=20000$. The scatterplot shows all observations 
within a slab perpendicular to the $z$ coordinate with $0.8\leq z\leq 1.2$. We plot the histograms and kernel density 
estimates along a slice where $z = 1$, and we only show rectangles where the empirical density is at least 
$2\times10^{-4}$ so that we do not plot empty rectangles. }
        \label{fig:mixture3d}
\end{figure}

We compare three ways to visualize the data.
First, we use a kernel density estimate. In more detail, we use a Gaussian kernel and select the
  bandwidth with biased cross-validation \cite{sain1994cv} in two dimensions and with 
  the plug-in bandwidth estimator \cite{sheather1991kernel, wand1994plugin, chacon2010plugin} in three dimensions. 
Both estimates are implemented in the R package \texttt{ks}. The kernel density estimates are shown in Figures 
\ref{fig:kerned2d} and \ref{fig:kerned3d}. 
In Figure \ref{fig:mixture3d} we plot the estimated density along the plane $z = 1$ as well as observations that lie in a 
slab where $0.8\leq z\leq 1.2$. 
    
Second, we plot a histogram with a fixed number of 15 equally sized bins in each dimension,
see Figures \ref{fig:fixed2d} and \ref{fig:fixed3d}. For the three-dimensional mixture 
the resulting histogram has $15^3=3375$ bins, of which only 834
are not empty. 
  
Third, we show the Beta-tree histogram with confidence level $1-\alpha = 90\%$ in Figures \ref{fig:hist2d} 
and \ref{fig:hist3d}. 
The Beta-tree histogram consists of only 25 rectangles in the two-dimensional setting and 125 rectangles in the 
three-dimensional setting. This exemplifies how the Beta-tree histogram yields a more parsimonious summary of the data 
compared to a histogram with a fixed number of bins. 

Figures \ref{fig:bounded2d} and \ref{fig:bounded3d} show the Beta-tree histogram with bounding box, which is obtained
by exluding 0.5\% of the data in each tail of each coordinate.
The Beta-tree histogram with bounding box consists of 36 rectangles in the 
two-dimensional setting and 315 rectangles in the three-dimensional setting.

We note that Figure~\ref{fig:mixture3d} suggests that all four methods are able to distinguish the first and
the second components in the mixture. However, only the two Beta-tree methods provide a confidence statement
to this effect. Section~\ref{section:mode-detection} gives more details for such multivariate mode hunting with
finite sample guarantees.

%Appendix \ref{appendix:additional} explores how varying the sample size and the confidence level $1-\alpha$ changes the
%number of rectangles in the Beta-tree.

%% file: simulation2_mode.tex
\section{Multivariate mode hunting}\label{section:mode-detection}

This section gives an example of how the Beta-tree  can be used for inference, namely to
perform multivariate mode detection with finite sample guarantees. The interest in multivariate mode
hunting derives from the important problem of detecting subpopulations in a distribution.
One prominent approach to this problem identifies such subpopulations with high density regions, see e.g.
\cite{Hartigan1975,GoodGaskins1980,Silverman1981, Minotte1997,FriedmanFisher1999,Ooi2002,burman2009mode,nugentstutzle}.
This gives rise to the problem of finding confidence bounds for the number and location of modes in a density.
Multivariate mode hunting is considered to be a difficult statistical problem due to the inherent multiple testing 
and due to the curse of dimensionality. Indeed, the statistical statements in the above references are typically
of approximate or asymptotic nature.

Here we show how mode hunting can be performed by using the Beta-tree as a summary of the data. The advantage of
using the Beta-tree for this task is that the analysis inherits the statistical guarantees that come with the
Beta-tree. 

In order to check that a density $f$ has two separate modes at locations $x$ and $y$, it is necessary to check
that on every (possible curved) path that connects $x$ and $y$, there is a point $z$ with $f(z) <\min (f(x),f(y))$.
This poses not only a statistical problem due to the simultaneous estimation of $f$, but also a computational problem
since all paths would need to be checked. This has motivated various approximations proposed in the literature, such as
checking only along convex combinations $z_{\alpha} = \alpha x + (1-\alpha) y$, $0\leq \alpha\leq 1$,
see \cite{burman2009mode}. A Beta-tree makes it possible to avoid this restriction: Since the Beta-tree segments
$\R^d$ into rectangles for which we are confident that $f$ is approximately constant, it is sufficient
to check all paths of adjacent\footnote{Since $R_i=\bigtimes_{p=1}^d (l_{ip},u_{ip})$, 
$R_i$ and $R_j$ are adjacent iff $[l_{ip},u_{ip}] \cap [l_{jp},u_{jp}] \neq \emptyset$ for all $p=1,\ldots,d$.}
 rectangles in the Beta-tree. This is typically a manageable computational
task since the number of rectangles in the Beta-tree is not large. 
For example, the group of rectangles in dark green in the top left and lower right corners in Figure~\ref{fig:hist2d} 
have higher density compared to the rectangles in between. Indeed, the underlying mixture distribution has two modes 
at $(1.5, 0.6)$ and $(2, -1.5)$.
Moreover, the Beta-tree provides simultaneous confidence bounds (\ref{3}) for the average density
${f}(R) = \int_R f(x)\mathrm{d} x/|R|$ (which equals $f$ on $R$ if $f$ is constant on $R$). This makes
it straightforward to check the condition ${f}(R_1) < \min({f}(R_2) , {f}(R_3) )$ by checking whether the 
upper confidence bound for ${f}(R_1)$ is smaller than the lower confidence bounds for ${f}(R_2)$ and ${f}(R_3)$.
Since these confidence bounds are simultaneous, any statement involving such inequalities along multiple
paths will inherit the finite sample confidence level $1-\alpha$. In particular, it is possible to claim the
existence of a certain number of modes with a finite sample guarantee.

We summarize our algorithm in Algorithm \ref{alg:mode}. In short, we first tag the rectangle with highest empirical 
density as a mode. Then, we iterate through the rectangles in the Beta-tree in descending order of their empirical density. 
For each rectangle $R_i$, we iterate through every path from $R_i$ to every mode that has been tagged so far, and if 
we find a path where
\textit{no} rectangle along the path has lower density than both endpoints, then $R_i$ will not be tagged as a mode.  

% algorithm
\begin{algorithm}[ht]
\caption{Multivariate mode hunting using the Beta-tree}\label{alg:mode}
\KwIn{A Beta-tree with $N$ rectangles $R_i$ and their empirical density $h_i$ given by (\ref{6}) and
the lower and upper confidence bounds $f_L(R_i)$ and $f_U(R_i)$ given by (\ref{3}). The list of modes $\mathcal{M}$ is empty.}
Compute the adjacency matrix $A\in\mathbb{Z}^{N\times N}$ such that $A_{i,j} = 1$ iff $R_i$ and $R_j$ are adjacent\;
Order the rectangles $R_i$ such that $h_1 \geq h_2 \geq  \ldots \geq h_N$\;
Tag the region with the highest density as a mode:
$\mathcal{M} = \{R_1\}$\;
\For{$i = 2, \ldots, N$}{
       \For{$m \in \mathcal{M}$ (i.e. iterate through the current modes in $\mathcal{M}$)}{
       \For{Every path connecting $R_i$ and $m$}{
       \uIf{There is no rectangle $R$ along the path such that
      $f_U(R) < \min(f_L(m), f_L(R_i))$}{$R_i$ is not a mode. Break the two inner for-loops and
          move on to $R_{i+1}$, i.e. $i\leftarrow i+1$}
             }
  
       }
         Add $R_i$ as a mode: $\mathcal{M} = \mathcal{M}\cup \{R_i\}$
 }
\KwOut{The list of modes $\mathcal{M}$} 
\end{algorithm}

\begin{figure}[h]
     \centering
     \begin{subfigure}[b]{0.4\textwidth}
         \centering
         \includegraphics[width=\textwidth]{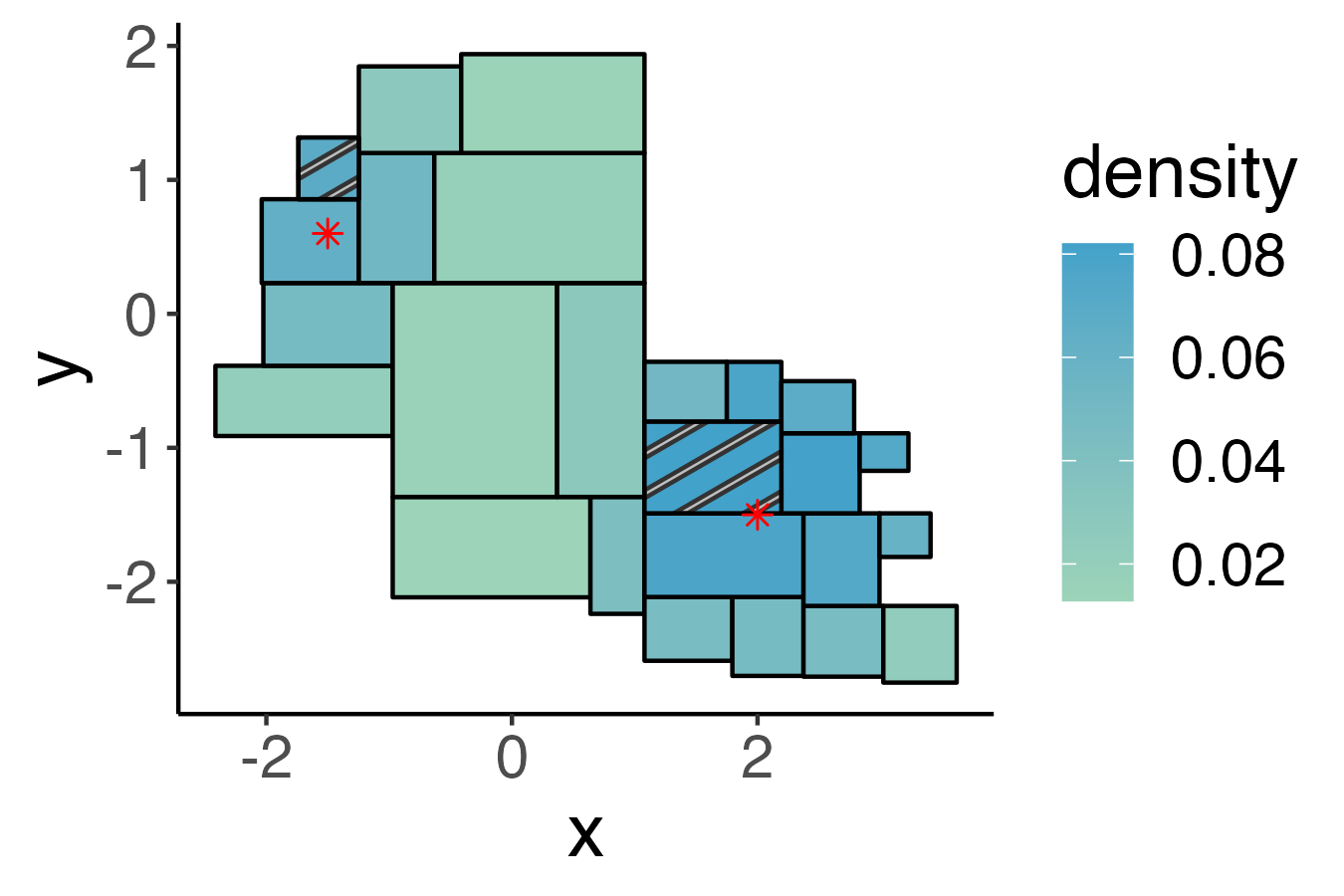}
       %  \caption{Kernel density estimate}
         \label{fig:mode}
     \end{subfigure}
     \hfill
     \begin{subfigure}[b]{0.4\textwidth}
         \centering
         \includegraphics[width=1\textwidth]{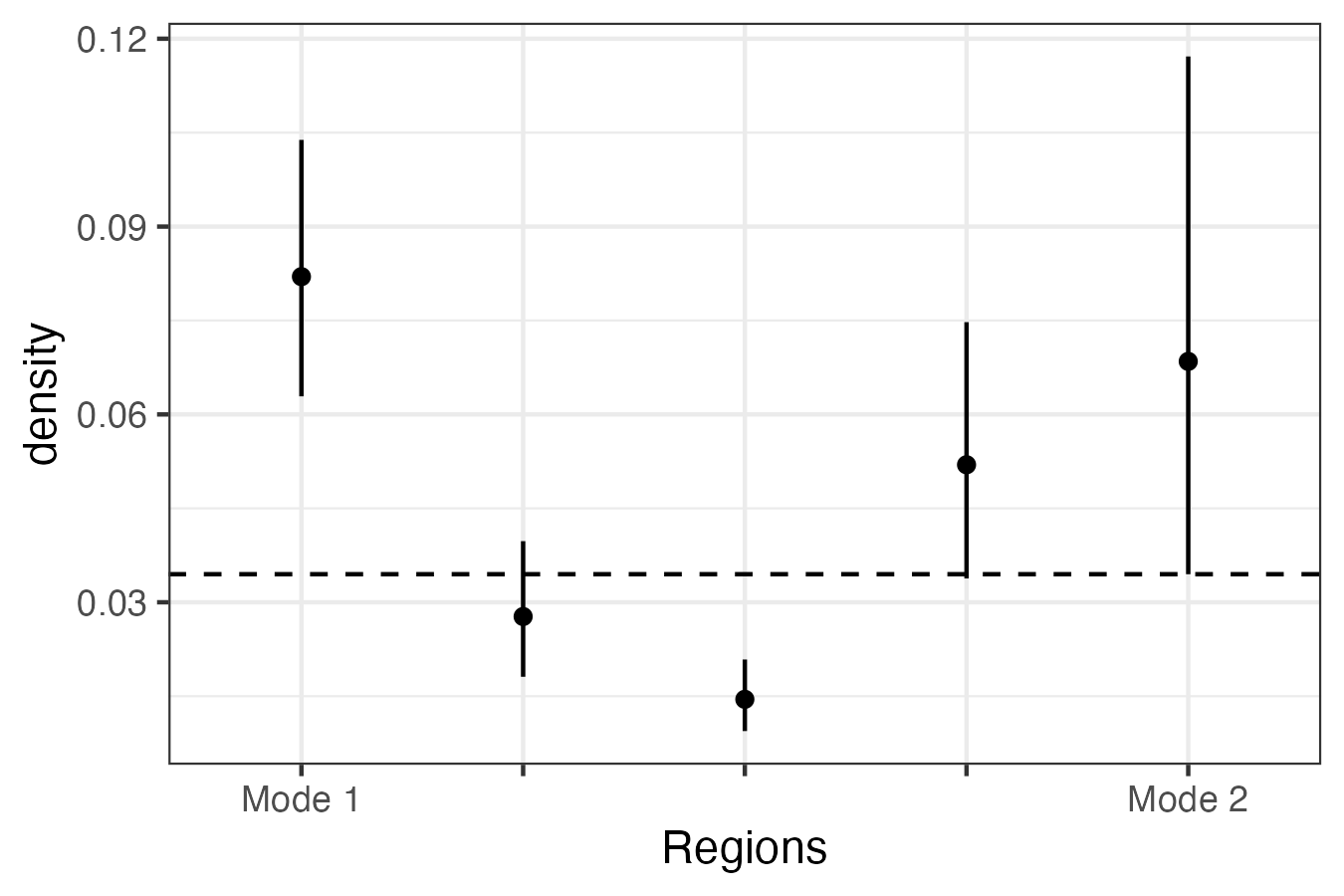}
        %  \caption{Histogram with 15 bins in each dimension}
         \label{fig:ci}
     \end{subfigure}
        \caption{Left: Beta-tree histogram of a two-dimensional Gaussian mixture with confidence level $1-\alpha = 0.9$. 
The modal rectangles are indicated by stripes and the red stars mark the two modes of the mixture distribution. 
Right: The confidence intervals for $f(R)$ for every rectangle along the shortest path between the two modes. 
The points indicate the empirical density in each rectangle and the dashed line shows the minimum of the 
lower confidence bounds for the two modal rectangles.}
        \label{fig:modefigs}
\end{figure}

We now apply our procedure to identify modes in the two mixture scenarios considered in Section \ref{section:visualization}.
For faster computation we only considered paths with
lengths at most 6.
%After we identify a list of candidate modes we can visually inspect if they are actually modes or not from the histogram). 
In the two-dimensional Gaussian mixture we identify two modes,
 whose corresponding rectangles are striped in
Figure~\ref{fig:modefigs} (left). The true modes (shown as red asterisks) are close to these two
rectangles. Figure~\ref{fig:modefigs} (right) shows the confidence intervals for the $f(R)$ along the shortest
path (in terms of number of rectangles) between the two identified modes. The plot shows that there exists a rectangle
whose upper confidence bound is below the minimum of the lower bounds for the two modal rectangles, 
which is marked by a dashed line.

For the three-dimensional mixture we are able to identify three modes. We
report the locations of true modes and the centers of the modal rectangles identified by the
algorithm in Table \ref{tab:mode}. The estimated modes are again close to the true modes. 
% (Note that when I use \(n=10,000\) observations I was only able to identify two modes out of three).

\begin{table}[h]
\caption{The coordinates of the centers of the modal rectangles identified by the algorithm (``Estimated modes'') and 
of the true modes. }\label{tab:mode}
\centering
\begin{tabular}{| c | c| c |}
\hline
Index & Estimated modes & True modes\\
\hline
1 & (-2.7,-3.0,-2.4) & (-2.6, -3, -2)  \\
2 & (-1.0,0.8,1.7) & (-1.5, 0.6, 1)\\
3 & (1.7,-1.6,0.3) & (2, -1.5, 0) \\
\hline
\end{tabular}
\end{table}

%% file: real.tex
\section{A real data example}\label{section:real}

We now apply our approach to visualize and identify cell populations in flow cytometry data. In \cite{brinkman2007fsc},
flow cytometry was used to analyze %biomarkers collected from
peripheral blood samples collected from patients who underwent a bone marrow
transplant. The objective was to
identify biomarkers which indicate graft-versus-host disease (GvHD), which occurs
in allogeneic hematopoietic stem cell transplant recipients when
donor-immune cells attack tissue of the recipient. Researchers
initially identified 121 subpopulations from 6 biomarkers, among which they pinpointed the
population identified as CD3+ CD4+ CD8b+ to have the highest correlation with the development of acute GvHD.

A subset of the data from this research is publicly available in \texttt{RvHD}
in the R package \texttt{mclust}. The data contain 9083 observations from a
patient with GvHD and 6809 observations from a control patient. The data includes
four biomarkers CD4, CD8b, CD3, and CD8. Since the sample size is
limited, we will construct a histogram using only the the first two variables CD4 and CD8b.

\begin{figure}
 \centering
     \begin{subfigure}[b]{0.45\textwidth}
         \centering
         \includegraphics[width=\textwidth]{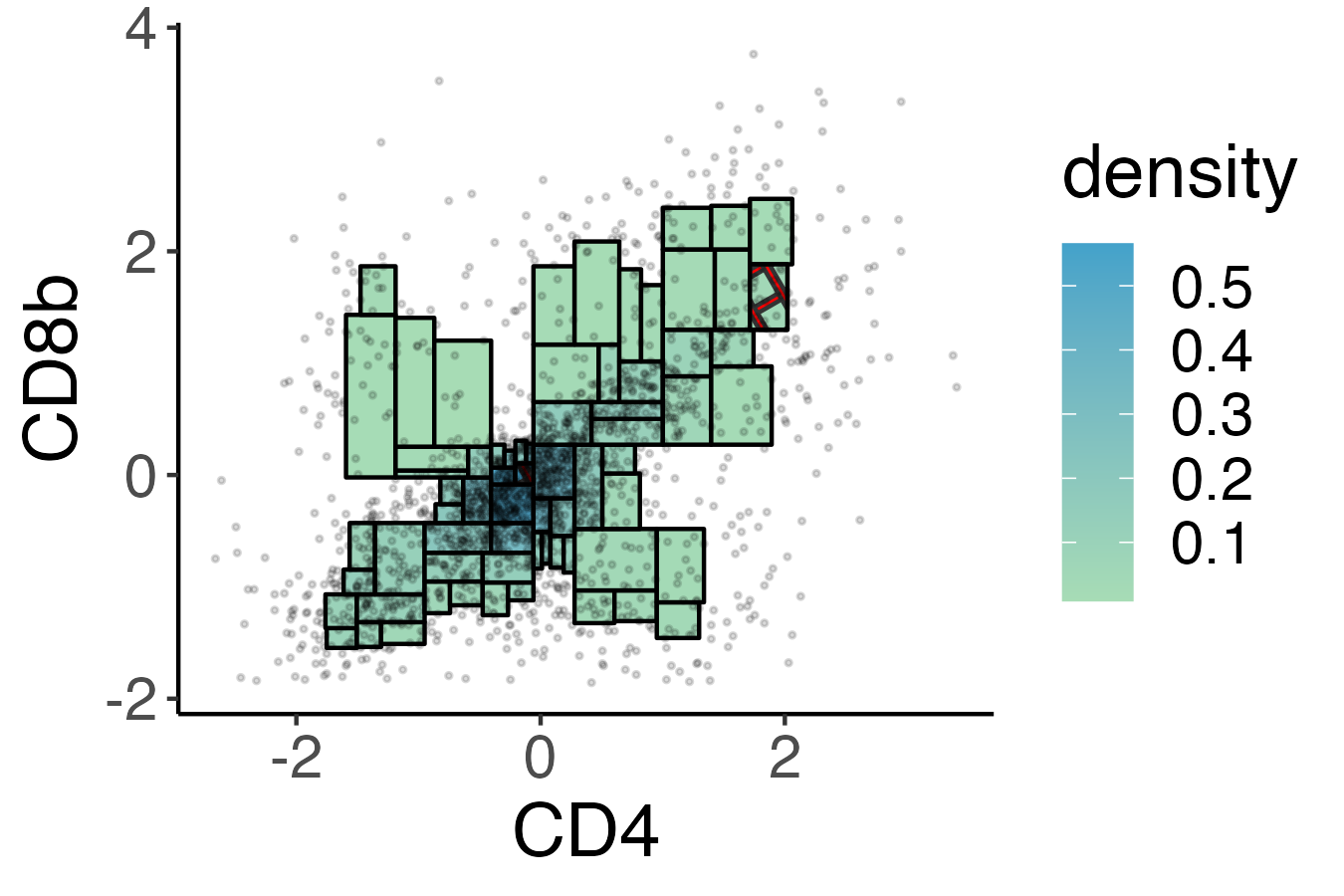}
         \caption{}
         \label{fig:real2d}
     \end{subfigure}
     \hfill
     \begin{subfigure}[b]{0.45\textwidth}
         \centering
         \includegraphics[width=1\textwidth]{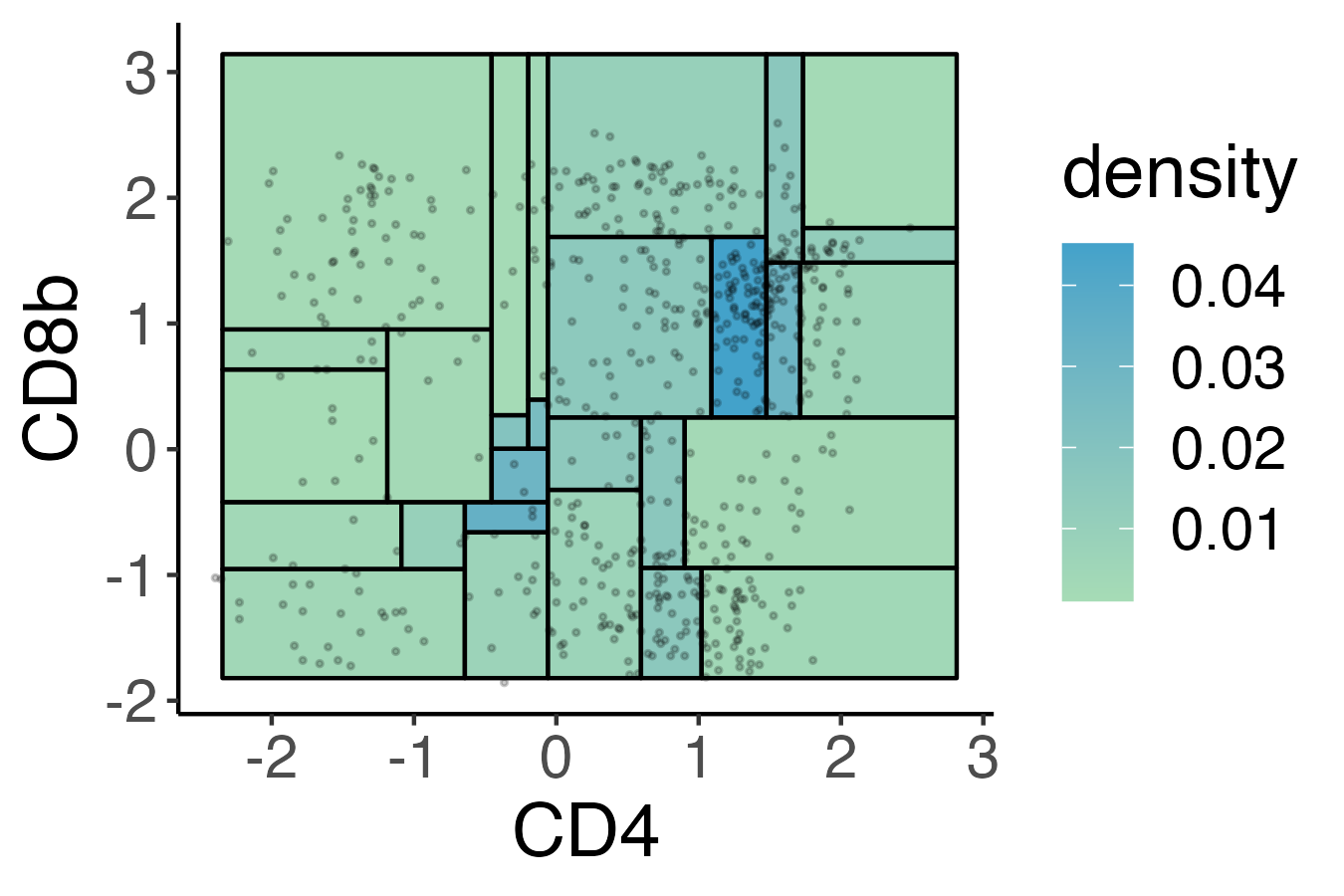}
        \caption{}
         \label{fig:real3dslice}
     \end{subfigure}
\label{fig:real}
\caption{Beta-tree histograms of data from the case patient, $1-\alpha = 0.9$. (a) A two-dimensional histogram of CD8b vs. CD4. 
The two identified modal rectangles are marked by red stripes. A random sample of 2000 observations is also shown. 
(b) A three-dimensional bounded Beta-tree histogram of CD4, CD8b, and CD3. The display shows rectangles in the Beta-tree 
histogram that intersect the plane $\mathrm{CD3}=1.0$. The displayed points are observations within the slab 
$0.8\leq \mathrm{CD3} \leq 1.2$.}
\end{figure}

We constructed a Beta-tree histogram using the data of the patient who developed GvHD (who we refer to as the case patient) 
and Algorithm~\ref{alg:mode} identifies two modes, which indicates the presence of two cell populations, 
see Figure \ref{fig:real2d}. We report the centers of the two 
modal rectangles in the column ``Center'' in Table \ref{tab:real}. If these two populations are
indicative of GvHD, then the empirical density in these regions from
the control patient, which is given in the column ``Density (control)'', should be lower
compared to that of the case patient. In this
example, the empirical density of the control patient
is indeed well below the confidence intervals for the case patient in both regions, suggesting that these regions might
be specific to GvHD. 

Finally, we compute a bounded Beta-tree histogram for CD4, CD8b, CD3 and visualize one slice of the histogram along the axis 
$CD3 = 1.0$. We also plot observations within the slab $0.8\leq CD3\leq 1.2$. The Beta-tree histogram captures one cluster of 
observations characterized by high values of CD4, CD8b and CD3, see Figure \ref{fig:real3dslice}.

\begin{table}[h]
\caption{Column ``Center'' shows the the coordinates of the center of the modal rectangles in the sample collected from 
the case patient. Column ``CI (Case)'' shows the confidence intervals for the average densities of these two modal regions. 
Column ``Density (Control)'' reports the empirical density of the data collected from the control patient. }\label{tab:real}
\centering
\begin{tabular}{| c | c| c |}
\hline
Center & CI (Case) & Density (Control) \\
\hline
(-0.13,0.01) & (0.42, 0.74) & 0.009 \\
(1.87, 1.59) & (0.03, 0.06) & 0.0012 \\
\hline
\end{tabular}
\end{table}

%% file: discussion.tex
\section{Discussion}

This paper introduces Beta-trees for summarizing multivariate data. The Beta-tree possesses several importantproperties: 
It provides a compact summary of the data by partitioning the sample space into rectangles on which the distribution is
close to uniform. The parition is adaptive to the data, which is key for avoiding the curse of dimensionality. The 
probability content of each rectangle in the partition has a sampling distribution that is known exactly and thus allows to set
finite sample confidence bounds. A multiscale Bonferroni adjustment results in simultaneous confidence bounds
whose widths do not depend on the dimension and match the optimal univariate widths, thus
avoiding the curse the dimensionality. The simultaneous validity of the confidence intervals allows to use Beta-trees
for various data-analytic tasks. As an example, we showed  how Beta-trees can be used for multivariate
mode hunting with finite sample guarantees. We illustrated this with flow cytometry data.

%% file: proof2.tex
\section{Proofs} \label{section:proof}

\subsection{Proof of Proposition~\ref{beta1}}

The proof of Proposition~\ref{beta1} is based on the following result:

\begin{Proposition}  \label{prop}
Let $X_1,\ldots,X_n \in \R^d$ be i.i.d. with distribution $F$ and let $g,h: \R^d \ra \R$
be measurable functions. Write $g(X)_{(k)}$ for the $k$th order statistic of the $g(X_i)$,
$i=1,\ldots,n$. For fixed $j$ and $k$ with $1 \leq j<k \leq n$ set
\begin{align*}
R\ &:=\ \Bl\{ x \in \R^d:\ g(x) < g(X)_{(k)} \Br\},\\
\{Y_1,\ldots,Y_{k-1}\}\ &:=\ \{X_1,\ldots,X_n\} \cap R,\\
S\ &:=\ \Bl\{ x \in R:\ h(x) < h(Y)_{(j)} \Br\}.
\end{align*}
Assume that $F,g,h$ are such that the $g(X_i)$ and the $h(Y_i)$ have a
continuous cdf, so $R$ contains $\#R=k-1$ observations a.s. and $\#S=j-1$ a.s.
Then
\begin{enumerate}
\item[(a)] $F(R) \sim \textrm{Beta } (\#R +1, n-\#R)$
\item[(b)] $\frac{F(S)}{F(R)} \sim \textrm{Beta } (\#S +1, \#R-\#S)$
\item[(c)] $\frac{F(S)}{F(R)}$ and $F(R)$ are independent
\item[(d)] $F(S) \sim \textrm{Beta } (\#S +1, n-\#S)$
\item[(e)] The results (a)--(d) continue to hold when the above construction is iterated,
starting with the above $S$ in place of $R$ and using prescribed functions
$g$ and $h$ that may be different from the initial functions.
\end{enumerate}
\end{Proposition}

We point out that this result requires that $k$ and $j$, and hence $\#R$ and
$\#S$, are prescribed (i.e. do not depend on the data), and that the inequalities in
the definition of $R$ and $S$ are strict as using `$\leq$' for $R$ will add
one observation to $R$ which will generally invalidate the beta distribution
for $F(S)$.

The first such result about the beta distribution of $F(S)$ for multivariate $S$
constructed from  order statistics appears to be Wald~(1943), who considered
the special case where $g$ and $h$ are univariate marginals \cite{Wald1943}. Wald's
proof has an important gap which was patched by Tukey~(1947)
with a lemma that he called `Wald's Principle' \cite{Tuckey1947}. The works of Wald and
Tukey are hampered by the methodology available in the 1940s. We provide
a short and elementary proof of Proposition\ref{prop} and a more general
version of Wald's Principle in Lemma~\ref{Wald}.

\begin{Lemma} \label{Wald}
(Wald's Principle)
Let $X_1,\ldots,X_n \in \R^d$ be i.i.d. with distribution $F$ and let $g: \R^d \ra \R$ be
a measurable function. Suppose $F$ and $g$ are such that the univariate random variables
$g(X_i)$ have a continuous cdf . Fix $k \in \{1,\ldots,n\}$ and write $g(X)_{(k)}$ for
the $k$th order statistic of the $g(X_i)$, $i=1,\ldots,n$. Then divide the $X_i$ into two
groups according to whether $g(X_i)$ is smaller or larger than $g(X)_{(k)}$:
\begin{align*}
\{Y_1,\ldots,Y_{k-1}\} & :=  \{X_i:\ g(X_i) < g(X)_{(k)}\} \\
\{Z_1,\ldots,Z_{n-k}\} & :=  \{X_i:\ g(X_i) > g(X)_{(k)}\}
\end{align*}
where the $Y_i$ and the $Z_i$ are enumerated in the original order of outcome
among the $X_i$.

Then, conditional on $g(X)_{(k)}=t$:

\begin{enumerate}
\item[(a)] The $\{Y_i\}$ and the $\{Z_i\}$ are independent.
\item[(b)] The $Y_i,\ldots Y_{k-1}$ are i.i.d. with distribution $G(\cdot)=\frac{F(\cdot\, \cap R_t)}{
F(R_t)}$ and the $Z_1,\ldots,Z_{n-k}$ are i.i.d. with distribution
$K(\cdot)=\frac{F(\cdot \,\cap R_t^c)}{F(R_t^c)}$, 
where $R_t:=\{x \in \R^d:\ g(x)<t\}$.
\end{enumerate}
\end{Lemma}

The lemma can be seen as a generalization of the following well known fact about
univariate order statistics: Conditional on $X_{(k)}=t$, the vectors $(X_{(1)},\ldots,X_{(k-1)})$
and $(X_{(k+1)},\ldots,X_{(n)})$ are independent and the joint law of $(X_{(1)},\ldots,X_{(k-1)})$
is that of the order statistics of $k-1$ i.i.d. random variables with distribution $\frac{F(\cdot \,\cap
(-\infty,t))}{F((-\infty,t))}$, see Theorem 1.8.1 in \cite{Reiss1989Order}. Lemma~\ref{Wald} generalizes this
result by establishing that the unordered $X_i$ corresponding to $(X_{(1)},\ldots,X_{(k-1)})$
are i.i.d., and by generalizing this result to multivariate $X_i$ ordered via $g$.
\bigskip

{\bf Proof of Lemma~\ref{Wald}:} We first note that $g(X_1)$ having a continuous
cdf implies that $g(X_i)=g(X)_{(k)}$ for exactly one index $i$ a.s., hence there are
a.s. $k-1$ observations $Y_i$ and $n-k$ observations $Z_i$.

For Borel sets $B_i,\tilde{B}_i \in \R^d$:

\begin{multline}
\Pr \Bl(Y_1 \in B_1,\ldots,Y_{k-1} \in B_{k-1},Z_1 \in \tilde{B}_1,\ldots,Z_{n-k} \in
\tilde{B}_{n-k} \Big| g(X)_{(k)} \in [t,t+dt)\Br)  \\
 = \frac{ \Pr \Bl({\cal A} :=\Bl\{Y_i \in B_i \cap R_t \mbox{ for } i=1,\ldots,k-1,
Z_j \in \tilde{B}_j \cap R_{t+dt}^c \mbox{ for } j=1,\ldots,n-k,g(X)_{(k)} \in [t,t+dt)\Br\}\Br)}{
\Pr \Bl( g(X)_{(k)} \in [t,t+dt) \Br)} \label{L1}
\end{multline}

We now write ${\cal A}$ in terms of the $X_i$: Let $T_{k,n}$ be the set of permutations $\tau$
of $\{1,\ldots,n\}$ such that $\tau(1) < \tau(2) < \ldots < \tau (k-1)$ and $\tau (k+1)<
\tau (k+2) < \ldots \tau (n)$. Since the $Y_i$ and the $Z_j$ are enumerated in the original
order of outcome among the $X_i$ we must have $Y_i=X_{\tau (i)}$ and $Z_j=X_{\tau(k+j)}$
for some $\tau \in T_{k,n}$. In fact, it is readily seen that ${\cal A} =\bigcup_{\tau \in T_{k,n}}
B_{\tau}$, where
$$
B_{\tau}\ :=\ \Bl\{X_{\tau (i)} \in B_i \cap R_t \mbox{ for } i=1,\ldots,k-1,\,
X_{\tau (k+j)} \in \tilde{B}_j \cap R_{t+dt}^c \mbox{ for } j=1,\ldots,n-k,\,
X_{\tau (k)} \in R_{t+dt} \setminus R_t \Br\}.
$$
The $B_{\tau}$ are mutually disjoint because different $\tau$ result in different assignments
of the $X_i$ to $R_t$, $R_{t+dt}\setminus R_t$ and $R_{t+dt}^c$. There are $\binom{n}{k-1}(n-k+1)
=\frac{n!}{(k-1)! (n-k)!}$ permutations in $T_{k,n}$: there are $\binom{n}{k-1}$ ways to choose
$\tau(1) < \ldots < \tau (k-1)$ and $n-k+1$ possibilities for $\tau (k)$, which then uniquely
determine $\tau (k+1) < \ldots < \tau (n)$. Therefore
\be \label{L2}
\Pr ({\cal A}) = \sum_{\tau \in T_{k,n}} \Pr (B_{\tau}) = \frac{n!}{(k-1)! (n-k)!} \left( \prod_{i=1}^{k-1}
F(B_i \cap R_t)\right) \left( \prod_{j=1}^{n-k} F(\tilde{B}_j \cap R_{t+dt}^c) \right)\ dF_g(t)
\ee
where $F_g$ denotes the cdf of $g(X_1)$ and $dF_g(t) =F_g(t+dt)-F_g(t)$. As for the denominator
of (\ref{L1}), since $F_g$ is continuous it is known that the univariate $k$th order statistic
$g(X)_{(k)}$ has the following density w.r.t. the cdf $F_g$:
$$
\Pr \Bl(g(X)_{(k)} \in [t,t+dt)\Br) =\frac{n!}{(k-1)! (n-k)!} \Bl(F_g(t)\Br)^{k-1} \Bl( 1-F_g(t)\Br)^{n-k}
dF_g(t),
$$
see Theorem 1.5.1 in \cite{Reiss1989Order}. Using $F_g(t)=F(R_t)$ and (\ref{L2}) shows that (\ref{L1}) equals
$$
\prod_{i=1}^{k-1} \frac{F(B_i \cap R_t)}{F(R_t)}\ \prod_{j=1}^{n-k} \frac{F(\tilde{B}_j \cap R_t^c)}{
F(R_t^c)}
$$
which establishes the claims of the lemma. 

As an aside we note that it appears to be surprisingly difficult to establish this result via 
conditional distributions
rather than by calculating probabilities of $[t,t+dt)$. In the univariate setting, Theorem 1.8.1
in \cite{Reiss1989Order} shows that conditional on $X_{(k)}$, the order statistics to the left of $X_{(k)}$
are independent of those to the right, and the proof of Corollary 1.8.4 and problem 1.33 can
be used to extend this result to the unordered observations as in claim (a) of the Lemma. 
Already in that univariate
setting these proofs with conditional distributions are rather complicated. $\Box$
\bigskip

{\bf Proof of Proposition~\ref{prop}:} We will use the following two well known facts:
\begin{Fact}  \label{factA}
If the univariate $Z_1,\ldots,Z_n$ are i.i.d. with a continuous cdf $G$, then
$G(Z_{(k)}) \sim \textrm{ Beta }(k, n+1-k)$.
\end{Fact}

\begin{Fact}  \label{factB}
If $V \sim \textrm{ Beta }(\alpha, \beta)$ and $W \sim \textrm{ Beta }(\alpha +\beta,
\gamma)$ are independent, the $VW \sim \textrm{ Beta }(\alpha, \beta +\gamma)$.
\end{Fact}

Write $F_g$ for the univariate cdf of $g(X_1)$. Then $F(R) =F_g\Bl(g(X)_{(k)}\Br)
\sim \textrm{ Beta }(k,n+1-k)$ by Fact~\ref{factA}, proving (a).

By Lemma~\ref{Wald}, conditional on $g(X)_{(k)}$ the $Y_1,\ldots,Y_{k-1}$ are i.i.d.
with distribution $G(\cdot)=\frac{F(\cdot\, \cap R)}{F(R)}$. Write $G_h$ for the cdf
of $h(Y_1)$, so for real $t$:
$$
G_h(t)\ =\ G\Bl(\{ x \in \R^d:\ h(x) \leq t \} \Br)\ =\ \frac{F\Bl(\{x \in \R^d: h(x) \leq t\}
\cap R\Br)}{F(R)}.
$$
By the definition of $S$:
$$
\frac{F(S)}{F(R)}\ =\ G_h\Bl(h(Y)_{(j)}\Br) \sim \textrm{ Beta }(j,k-j)
$$
by Fact~\ref{factA}. Since this conditional distribution given $g(X)_{(k)}$, i.e. given $R$,
does not depend on $R$ as $\#R=k-1$ is fixed, it follows that this result also holds
unconditionally and that $\frac{F(S)}{F(R)}$ and $F(R)$ are independent, proving (b) and (c).

(d) follows from Fact~\ref{factB} and (a)--(c): set $\alpha:=j$, $\beta:=k-j$, $\gamma :=
n+1-k$ in Fact~\ref{factB}.

As for (e), the above proof goes through if one iterates this construction starting with
$S$ in place of $R$. In particular, applying Lemma~\ref{Wald} again to the conditional
distribution given $R$, $G(\cdot)=\frac{F(\cdot\, \cap R)}{F(R)}$, shows that the conditional
distribution given $S$ is
$$
\frac{\frac{F(\cdot \, R \cap S)}{F(R)}}{\frac{F(S \cap R)}{F(R)}}\ =\ \frac{F(\cdot \, \cap S)}{
F(S)}\ \ \ \mbox{ since $S \subset R$ }. \ \Box
$$

{\bf Proof of Proposition~\ref{beta1}}: This follows from Proposition~\ref{prop} by taking
$g,h$ to be functions of the form $x \mapsto \pm x_p$ for $p \in \{1,\ldots,d\}$,
i.e. by selecting a certain univariate marginal and choosing the sign of $x_p$
to select the observations to the left or to the right of the order statistic of that
marginal. Note that $F$ being continuous implies that these $g(X_i)$ and $h(Y_i)$
have a continuous cdf. This proposition also applies to selecting observations inside a
bounding box if that box is constructed as described in Section~\ref{section:construct_kd}. $\Box$ 
\bigskip

\subsection{Proof of Theorem~\ref{optimal}}

Consider $R_k$ at tree level $D$ and write $p_D:=\frac{n_k+1}{n+1}$. Then Proposition~\ref{beta1} gives
$F(R_k) \sim \textrm{ Beta }\left((n+1)p_D,(n+1)(1-p_D)\right)$. 
Let $x_{up}:=q\textrm{Beta }\left(1-\frac{\alp_D}{2},(n+1)p_D,(n+1)(1-p_D)\right)$
be the upper confidence bound of $C_k(\alp_D)$. Then
$$
\frac{\alp_D}{2}\ =\ \Pr \left(F(R_k) \geq x_{up}\right)\ \leq \ \exp \Bl( -(n+1) \Psi(x_{up},p_D)\Br)
$$
by Proposition~2.1 in \cite{Du98}, where $\Psi(x,p):=p \log \frac{p}{x} +(1-p) \log \frac{1-p}{1-x}$. Hence
$\Psi(x_{up},p_D) \leq \frac{1}{n+1} \log \frac{2}{\alp_D}$. Using the inequality at the end of said proposition, this implies
$$
x_{up}-p_D\ \leq \ \sqrt{2p_D(1-p_D) \frac{\log \frac{2}{\alp_D}}{n+1}} +(1-2p_D)^+ \frac{\log \frac{2}{\alp_D}}{n+1}
$$
In the same way one finds an (even tighter) inequality for the lower confidence bound $x_{low}$ of $C_k(\alp_D)$:
$$
x_{low}-p_D\ \geq \ - \sqrt{2p_D(1-p_D) \frac{\log \frac{2}{\alp_D}}{n+1}}
$$
Therefore
\be \label{optimalst}
\sup_{x \in C_k(\alp_D)} 
\sqrt{n} \frac{|x-p_D|}{\sqrt{p_D(1-p_D)}}\ \leq \ \sqrt{2 \log \frac{2}{\alp_D}} +\frac{\log \frac{2}{\alp_D}}{
\sqrt{np_D(1-p_D)}}
\ee
Using $2^D \sim \frac{1}{p_D}$, $D_{max} \sim \log_2 n$ and $\sum_{B=2}^{D_{max}} \frac{1}{B}
\sim \log \log_2 n$, we obtain $\frac{2}{\alp_D} \leq 2\frac{(\log_2n) (\log \log_2 n)}{\alp \,p_D}$, hence
$$
\log \frac{2}{\alp_D} \leq (1+\eps_n) \log \frac{e}{p_D},\qquad \mbox{with }\eps_n:= \frac{\log \Bl( \frac{2}{\alp} (\log_2 n)
(\log \log_2 n)\Br)}{\log n^{1-q}}
$$
since $p_D \leq n^{q-1}$. Furthermore, $p_D \geq \frac{\log^2 n}{n}$ yields
$$
\frac{\log \frac{2}{\alp}}{\sqrt{n p_D (1-p_D)}} \leq \frac{\frac{3}{2} \log \frac{e}{p_D}}{\log n}
\leq \frac{3}{2} \sqrt{\frac{\log \frac{e}{p_D}}{\log n}}
$$
This allows to bound (\ref{optimalst}) as follows:
\begin{align}
\sup_{x \in C_k(\alp_D)} \sqrt{n} \frac{|x-p_D|}{\sqrt{p_D(1-p_D)}} &\leq \sqrt{2(1+\eps_n) \log \frac{e}{p_D}}
+ \frac{3}{2} \sqrt{\frac{\log \frac{e}{p_D}}{\log n}} \nn \\
&\leq \Bl( \sqrt{2} +\eps_n +\frac{3}{2} \sqrt{\frac{1}{\log n}} \Br) \sqrt{\log \frac{e}{p_D}} \label{optimalstst}
\end{align}
This essentially establishes the claim apart from having $p_D$ in place of $x$ in the denominator. However,
(\ref{optimalstst}) yields $\sup_{x \in C_k(\alp_D)} |x-p_D| \leq \sqrt{\frac{p_D}{n}} \sqrt{3 \log \frac{e}{p_D}}$, so
$$
\sup_{x \in C_k(\alp_D)} \left| \frac{x}{p_D}-1\right| \ \leq \ \sqrt{\frac{3 \log \frac{e}{p_D}}{np_D}} \ \leq \ 
\sqrt{\frac{3}{\log n}}
$$
which gives $\sup_{x \in C_k(\alp_D)} \sqrt{\frac{p_D(1-p_D)}{x(1-x)}} \leq 1+\sqrt{\frac{3}{\log n}}$. Therefore
\begin{align*}
\sup_{x \in C_k(\alp_D)} \sqrt{n} \frac{|x-p_D|}{\sqrt{x(1-x)}} 
&\leq \Bl(\sqrt{2}+\eps_n +\frac{3}{2\sqrt{\log n}}\Br)\left(1+\sqrt{\frac{3}{\log n}}\right) \sqrt{\log \frac{e}{p_D}}\\
&\leq \Bl(\sqrt{2}+\frac{4}{\sqrt{\log n}}\Br) \sqrt{\log \frac{e}{p_D}}
\end{align*}
for $n$ large enough. $\Box$

%% file: appendix_kdtree.tex
\section{More details about $k$-d trees and Beta-trees}\label{appendix:kdtree}

This section gives some technical details about $k$-d trees that are relevant
for implementing Beta-trees as a data structure.

The indexing $R_k$ given in Section~\ref{section:construct_kd} is helpful if one wants 
to use algorithms that are not recursive.
The standard definition of a $k$-d trees identifies a node with a splitting hyperplane.
For implementing and working with a Beta-tree it is helpful to let a node represent a rectangle $R_k$
and then add the following information
to each node: The specification of $R_k$ in terms of two bounds (which may be infinite) for each coordinate,
the tree depth of the node (with the root node $R_0$ having depth 0), whether $R_k$ is a leaf
or not, 
the number $n_k$ of data in the interior of $R_k$, lower
and upper confidence bounds for $f(R_k)$ given in \eqref{3} and \eqref{eq:ci_tilde}, and the pointers to the two children nodes.

For example, if the root node represents $R_0=\R^d$, then $R_0$ contains
all the data, so $n_0=n$. 
The children have depth 1 and $n_1=\lceil \frac{n_0}{2} \rceil-1$, $n_2=n_0-\lceil \frac{n_0}{2} \rceil$.
More generally, one obtains for $k\geq 0$: $n_{2k+1}=\lceil \frac{n_k}{2} \rceil-1$, $n_{2k+2}=n_k-\lceil \frac{n_k}{2} \rceil$.
If one uses a bounding box as described in Section~\ref{section:construct_kd}, then $R_0$ is the bounding box 
and $n_0$ is the number of observations in the bounding box rather than $n$.

At tree depth $D\geq 0$ the $k$-d tree partitions $R_0$ into $2^D$ rectangles if all rectangles are split (which
may not be the case for larger $D$ due to the stopping criterion and the fact that the rectangles at a given
depth don't contain exactly the same number of observations.)
If $R_0$ is a bounding box, then
all of these rectangles are bounded, hence $N_D=2^D$ if all rectangles are split, whereas if $R_0=\R^d$ and
if all rectangles are split, then the number of bounded rectangles can be shown to be 
$N_D =  \prod_{i=1}^{d}\left(2^{\lceil \frac{D+1-i}{d}\rceil} - 2\right)_+$, 
which is zero for $D<2d$ but nonzero for $D \geq 2d$. Therefore in the case $R_0=\R^d$ 
the weighted Bonferroni adjustment
(\ref{eq:weighted_bonferroni}) needs to be modified as follows:
$$
\alpha_D = \frac{\alp}{N_D (D_{max}-D+2) \sum_{B=2}^{D_{max}-2d+2} \frac{1}{B}},\ \ D\geq 2d,
$$
and $\alpha_D=0$ for $D<2d$.
$N_D$ can be easily found by traversing the $k$-d tree. This is preferrable to the formulas for $N_D$ given above
since those hold only as long as all rectangles at a given depth are split.

Assuming all $n$ observations are distinct in each of the $d$ coordinates (which is the case with
probability 1 if the distribution $F$ is continuous), then each splitting hyperplane contains exactly one observation $X_i$.
Therefore at tree depth $D$ there are $2^D$ rectangles $R_k$ and $2^D-1$ observations $X_i$ that
don't belong to any of those $R_k$ since they lie on the boundary of some $R_k$. This motivates
to define the empirical measure as $F_n(R_k)=\frac{n_k+1}{n}$ rather than $F_n(R_k)=\frac{n_k}{n}$. This convention
also turns out to be convenient for formulating the technical results in the paper.

{\bf Finding the maximal rectangles that are bounded and satisfy the goodness-of-fit (GOF) condition}: \\
\nin Due to the tree
structure, this can be done recursively as follows:

Starting with the rectangle $R$ being the root node: If $R$ is bounded and satisfies the GOF condition
$\widetilde{\mathrm{lower}}(R) \leq h \leq \widetilde{\mathrm{upper}}(R)$, then $R$ is a
maximal rectangle. Else if $R$ is not a leaf, then call this procedure for the first child of $R$ in place of $R$, then
for the second child of $R$ in place of $R$.
\smallskip

\nin Alternatively, an iterative algorithm proceeds as follows:

{\bf for} tree depths $D=0,\ldots,D_{max}$:\\
\hspace*{1cm} {\bf for} rectangles $R$ at tree depth $D$:\\
\hspace*{1.5cm} {\bf if} ($R$ is bounded and satisfies the GOF condition) {\bf then} \  ($R$ is maximal; delete the subtree below $R$)